\documentclass[sigplan,nonacm]{acmart}
\usepackage{hyperref}
\usepackage{gensymb}
\usepackage[normalem]{ulem}
\usepackage{xcolor}
\usepackage{amsmath,amsfonts}
\usepackage{algorithmic}
\usepackage{graphicx}
\usepackage{textcomp}
\usepackage{enumitem}
\usepackage{listings}
\usepackage{subcaption}
\usepackage{soul}
\usepackage{comment}
\usepackage{tikz}
\usepackage{xspace}
\usepackage{array}
\usepackage{colortbl}
\usepackage{multirow}
\usepackage{makecell}
\usepackage{comment}
\usepackage{booktabs}
\hypersetup{
    unicode=true,
    bookmarksnumbered=true,
    colorlinks=true,
    linkcolor={red!70!black},
    citecolor={red!70!black},
    urlcolor={blue!70!black},
    pdfborder={0 0 0}
}
\usepackage[capitalize,nameinlink]{cleveref}
\usepackage{flushend}
\usepackage{pifont}

\lstdefinestyle{py-acm}{
  language=Python,
  basicstyle=\ttfamily\scriptsize,
  keywordstyle=\bfseries\color{blue!60!black},
  commentstyle=\itshape\color{green!40!black},
  stringstyle=\color{red!55!black},
  numberstyle=\tiny\color{gray},
  backgroundcolor=\color{gray!5},
  showstringspaces=false,
  columns=fullflexible,
  keepspaces=true,
  numbers=left,
  stepnumber=1,
  numbersep=6pt,
  tabsize=2,
  frame=single,
  framerule=0.3pt,
  breaklines=true,
  breakatwhitespace=true,
  captionpos=b,
  aboveskip=0.5em,
  belowskip=0.3em
}

\definecolor{gpuHigh}{RGB}{255,180,80}   % dark orange
\definecolor{gpuMed}{RGB}{255,210,130}   % medium orange
\definecolor{gpuLow}{RGB}{255,240,200}   % light orange

\definecolor{hgpuHigh}{RGB}{80,120,255}  % dark blue
\definecolor{hgpuMed}{RGB}{130,170,255}  % medium blue
\definecolor{hgpuLow}{RGB}{200,220,255}  % light blue

\definecolor{grayHigh}{RGB}{160,160,160}   % dark gray
\definecolor{grayMed}{RGB}{200,200,200}    % medium gray
\definecolor{grayLow}{RGB}{235,235,235}    % light gray

\definecolor{energyHigh}{RGB}{0,120,0}
\definecolor{energyMed}{RGB}{120,200,120}
\definecolor{energyLow}{RGB}{200,240,200}

\definecolor{costHigh}{RGB}{180,0,0}
\definecolor{costMed}{RGB}{240,120,120}
\definecolor{costLow}{RGB}{255,200,200}

\newcommand{\sysname}{Murakkab\xspace}
\newcommand{\sysnameshort}{Mrkb\xspace}

\newcommand{\myparagraph}[1]{\vspace{\smallskipamount}\noindent\textbf{#1.\xspace}}

\newcommand{\myparagraphemph}[1]{\vspace{\smallskipamount}\noindent\emph{#1.\xspace}}
\newcommand{\eg}{\emph{e.g.,}\xspace}
\newcommand{\etc}{etc.\@\xspace}

\newcommand{\ie}{\emph{i.e.,}\xspace}
\newcommand{\vs}{\emph{vs.}\xspace}

\newcommand*{\rom}[1]{\uppercase\expandafter{\romannumeral #1\relax}}

\newcounter{insightcounter}

\newif\ifshowcomment
\showcommenttrue
\ifshowcomment
    \newcommand{\gohar}[1]{{\color{blue}[GI: #1]}}
    \newcommand{\esha}[1]{{\color{purple}[EC: #1]}}
    \newcommand{\inigo}[1]{{\color{green}[IG: #1]}}
    \newcommand{\haoran}[1]{{\color{orange}[HQ: #1]}}
    \newcommand{\adam}[1]{{\color{lilac}[AB: #1]}}
    \newcommand{\rodrigo}[1]{{\color{magenta}[RF: #1]}}
    \newcommand{\todo}[1]{{\color{red}[TODO: #1]}}
\else
    \newcommand{\gohar}[1]{\ignorespaces}
    \newcommand{\esha}[1]{\ignorespaces}
    \newcommand{\inigo}[1]{\ignorespaces}
    \newcommand{\haoran}[1]{\ignorespaces}
    \newcommand{\adam}[1]{\ignorespaces}
    \newcommand{\rodrigo}[1]{\ignorespaces}
    \newcommand{\todo}[1]{\ignorespaces}
\fi

\usepackage[most]{tcolorbox}
\tcbset{textmarker/.style={%
        enhanced,
        parbox=false,boxrule=0mm,boxsep=0mm,arc=1.5mm,
        outer arc=1.5mm,left=2mm,right=2mm,top=4pt,bottom=3pt,
        toptitle=1mm,bottomtitle=1mm,oversize}}

\newtcolorbox{simplenoteBox}{colback=white, colframe=black, boxrule=0.2mm, arc=0mm, auto outer arc, boxsep=0mm, left=2mm, right=2mm, top=1mm, bottom=1mm} 
\newtcolorbox{noteBox}{textmarker,
    colback=gray!8!white}
\newcommand{\simplebox}[1]{\begin{simplenoteBox} #1 \end{simplenoteBox}}

\crefname{insight}{Insight}{Insight}
\newcounter{insight}
\newcommand{\boxinsight}[1]{%
  \refstepcounter{insight}%
  \vspace{\smallskipamount}%
  \noindent \simplebox{
    \textbf{\uline{Insight~\theinsight:}} 
    \textit{#1}% \par
  }
}

\title{\sysname{}: Resource-Efficient Agentic Workflow Orchestration in Cloud Platforms}

\author{
  Gohar~Irfan~Chaudhry$^{1}$, 
  Esha~Choukse$^{2}$, 
  Haoran~Qiu$^{2}$, 
  Íñigo~Goiri$^{2}$,\\
  Rodrigo~Fonseca$^{2}$, 
  Adam~Belay$^{1}$,
  Ricardo~Bianchini$^{3}$
}

\affiliation{%
  $^{1}$MIT CSAIL \hspace{1cm} $^{2}$Microsoft Azure Research -- Systems \hspace{1cm} $^{3}$Microsoft Azure
}

\begin{document}

\begin{abstract}
Agentic workflows commonly coordinate multiple models and tools with complex control logic. They are quickly becoming the dominant paradigm for AI applications.
However, serving them remains inefficient with today's frameworks.
The key problem is that they expose workflows as opaque sequences of model and tool calls that tightly couple agent logic with model and hardware choices.
Often, these workflow components are fragmented across different entities, preventing systems from reasoning about trade-offs across accuracy, latency, energy, and cost.
This leads to resource waste and degraded service-level objectives (SLOs).

We present \sysname{}, a resource-efficient serving system for agentic workflows. 
\sysname{} introduces a declarative abstraction that decouples workflow specification from execution configuration.
A profile-guided optimizer and adaptive runtime jointly manage the full stack: orchestrating workflow components, mapping them to models and hardware, and dynamically reconfiguring execution to satisfy user-defined SLOs.
By exposing the internal structure of agentic workflows, \sysname{} enables cross-layer optimization that existing frameworks and cloud schedulers cannot achieve.

Our evaluation on diverse workflows shows that \sysname{} reduces GPU usage by up to 2.8$\times$, energy consumption by 3.7$\times$, and cost by 4.3$\times$ while maintaining SLOs.
\end{abstract}

\maketitle

\pagestyle{plain}

\section{Introduction}

\begin{figure}
    \includegraphics[width=0.49\textwidth]{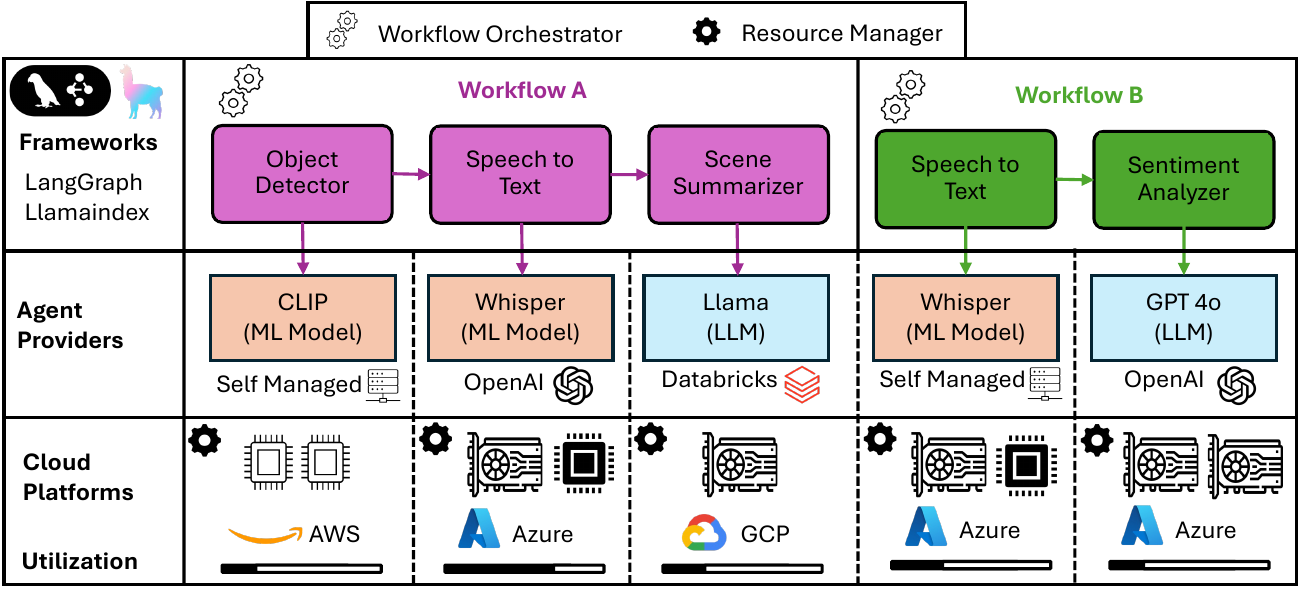}
    \caption{Today workflow developers use frameworks to call agents from \emph{different} providers hosted on \emph{multiple} cloud platforms. This fragmentation results in inefficiencies.}
    \label{fig:stack-today}
\end{figure}

Recent advancements in Large Language Models (LLMs) and multi-modal LLMs are rapidly reshaping fields like education~\cite{dai2024agent4edu,chu2025llm}, software engineering~\cite{jin2024llms,he2025llm}, and healthcare~\cite{goyal2024healai,cascella2023evaluating}.
Recent efforts have focused on extending these models with the ability to invoke external tools (\eg{} web browser or code execution) at inference time~\cite{wang2024gta,qin2023toolllm,zhang2024pybench,azure-ai-foundry-tools}, enabling them to take actions beyond static text generation.
This has paved the way for \textit{\textbf{agentic workflows}}, where multiple models and external tools collaborate to complete complex tasks, marking a shift from single-model inference to more sophisticated, compound AI applications.

Despite their promise, today’s agentic workflows are fragmented across agent frameworks, agent providers, and cloud platforms, often within different organizational boundaries (\Cref{fig:stack-today}):
developers use frameworks such as LangChain~\cite{langchain} or LlamaIndex~\cite{llamaindex} to stitch together models and tools, invoke provider APIs like OpenAI~\cite{openai_llms} or Databricks~\cite{databricks_llm_serving}, and rely on cloud providers for compute infrastructure.
Each layer pursues different goals (latency, quality, cost, or utilization), but coordination is minimal, leading to inefficiencies:
\begin{enumerate}[leftmargin=*]
\item \textbf{Tight coupling:} Hard-coded parameters (\eg{} models) and hardware choices hinder automated optimization.
\item \textbf{Disjoint orchestration:} Frameworks (construct workflows) and resource managers (deploy and serve workflows) operate in silos, causing suboptimal scheduling.
\item \textbf{Difficult trade-offs:} Accuracy, latency, energy, and cost objectives require navigating a large configuration space that grows with workflow depth and model/tool choices.
\end{enumerate}

These limitations drive up costs, degrade service-level objectives (SLOs), and waste resources.
Static, imperative definitions further make workflows brittle:
updating models or hardware requires manual refactoring and redeployment.

\begin{figure*}
    \centering
    \begin{subfigure}[b]{0.49\textwidth}
        \includegraphics[width=\textwidth]{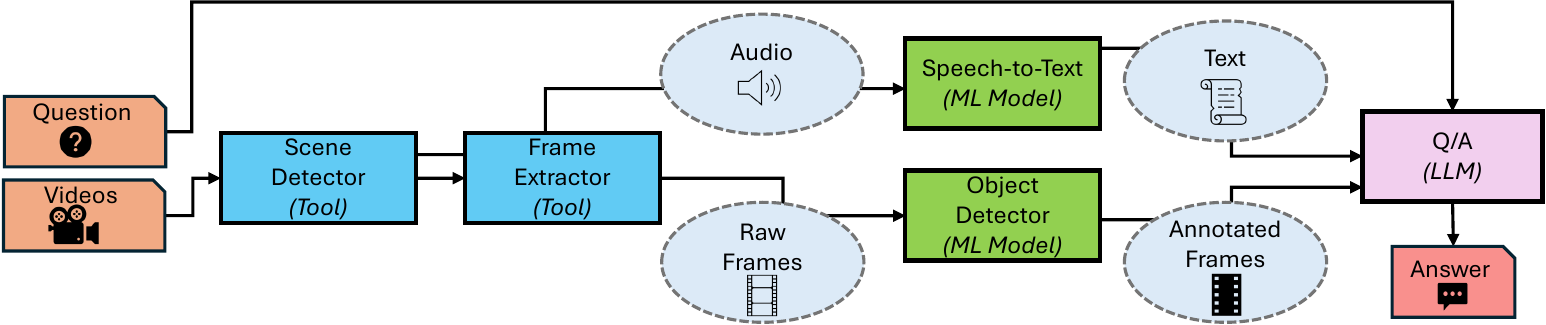}
        \caption{Video Q/A workflow: multi-modal with tools handling different modalities before feeding into an LLM for answering the query.}
        \vspace{-8pt}
        \label{fig:video-workflow}
    \end{subfigure}\hfill
    \begin{subfigure}[b]{0.49\textwidth}
        \includegraphics[width=\textwidth]{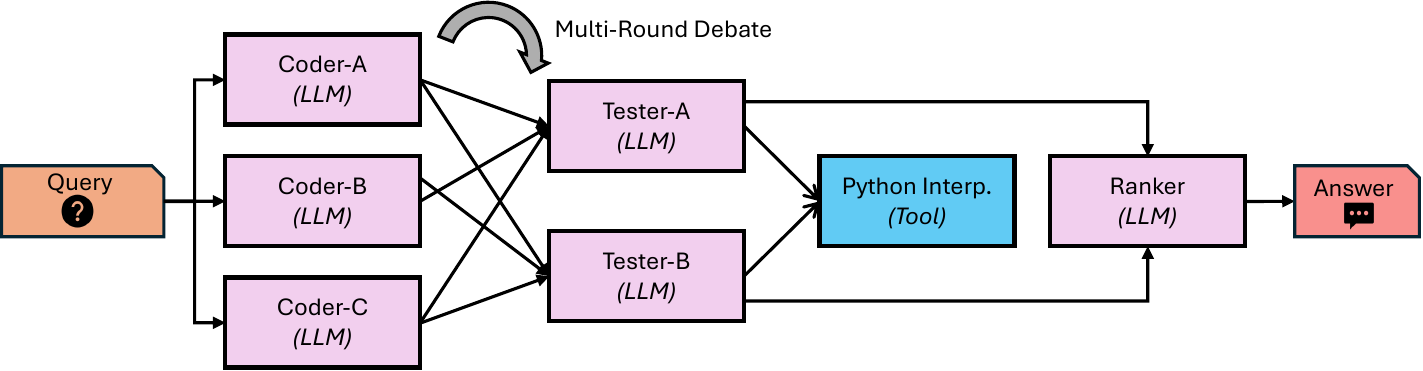}
        \caption{Code generation workflow: text-only with an LLM Debate structure to write, test and execute code.}
        \vspace{-8pt}
        \label{fig:code-gen-workflow}
    \end{subfigure}\hfill
    \caption{Two agentic workflows with different characteristics and components.}
    \label{fig:example-workflows}
\end{figure*}

\myparagraph{Our Work}
We present \sysname{}, a resource-efficient system for serving multi-tenant agentic workflows, built on two principles:
(1) declarative workflow specification and,
(2) adaptive, SLO-aware runtime system design.

Developers describe workflows with a declarative specification as \emph{logical} tasks and dependencies, decoupled from model, tool, or hardware choices.
This separation enables \sysname{} to integrate workflow orchestration with resource management, dynamically reconfiguring workflow parameters (\eg{} models or tools) and hardware configurations as needed, based on offline profiles and online monitoring to optimize quality, latency, energy, and cost objectives.

This design enables \sysname{} to efficiently explore the configuration space of workflow- and hardware-level knobs while meeting user-defined SLOs.
Unlike existing deployments that treat agentic workflows as black boxes, \sysname{} leverages end-to-end visibility to coordinate scheduling, colocation, and multiplexing across multi-tenant workflows.

We evaluate \sysname{} on representative workflows (video question answering, code generation and mathematical problem solving) using production-scale traces.
\sysname{} achieves up to 2.8$\times$ lower GPU usage, 3.7$\times$ less energy, and 4.3$\times$ lower cost than state-of-the-art baselines like LangGraph while preserving quality and latency SLOs.

\myparagraph{Summary}
This paper makes the following contributions:
\begin{itemize}[leftmargin=*]
\item Characterization of inefficiencies from workflow-, agent-, and hardware-level configurations. 
\item A declarative programming model and adaptive runtime unifying orchestration with resource management.
\item Profile-guided optimization and workflow-aware scheduling for multi-tenant workloads.
\item Evaluation showing significant efficiency gains without workflow quality or execution latency SLO violations.
\end{itemize}

\section{Background and Motivation}\label{sec:bg}

\subsection{Agentic Workflows}
We define an \emph{\textbf{agent}} as a composable unit that autonomously makes decisions, invokes tools, and participates in workflows, either independently or in collaboration with other agents~\cite{azure-agent-definition,google-agent-def}.
Each agent is typically powered by:
(1) a \emph{model} (\eg{} an LLM) for reasoning and language understanding,
(2) a set of \emph{instructions} specifying its goals, behavior, and constraints, and
(3) \emph{tools} for retrieving knowledge or taking actions.
When multiple agents (each with specialized roles) interact and collaborate to achieve complex goals, we call the resulting process an \emph{\textbf{agentic workflow}}.

Serving agentic workflows differs fundamentally from single-model inference.
Such workflows require coordinated interaction and data exchange among multiple agents to produce a final response.
These agents often span diverse use cases, modalities (\eg{} text, audio, images \etc{}), execution times, and software/hardware requirements.
Composing effective workflows thus demands a nuanced understanding of the trade-offs between task accuracy, request-serving latency, energy consumption, and hardware cost~\cite{kim2025costdynamicreasoningdemystifying}.

\subsection{Example Workflows}\label{sec:bg:example-workflows}
We consider three widely used, representative workflows~\cite{benchmarks}: \emph{Video Q/A}, \emph{Code Generation}, and \emph{Math Q/A} which exhibit distinct characteristics.
For the rest of the paper, we focus on the first two, as shown in \Cref{fig:example-workflows}, and include evaluation of the third one in \Cref{sec:appendix:math}.
A wide range of other workflows can be orchestrated using a similar approach.

\myparagraph{Video Q/A}
Video Q/A workflows are deployed for interactive query processing~\cite{viva2022romero}, security footage analysis~\cite{plixai2025}, and many other use-cases~\cite{tang2025videounderstandinglargelanguage,romero2021llamaheterogeneousserverless}.
We construct a comprehensive agentic workflow to capture these diverse use-cases~\cite{zhang2024omagent,agrawal2024agentic}.
This multi-modal workflow answers textual queries about input videos (\Cref{fig:video-workflow}).
Several agents in this workflow collaborate to produce the final result:
\begin{enumerate}[leftmargin=*]
    \item \emph{Scene Detector} to separate the input videos into distinct scenes that ensure better capture of key-frames and extract the audio clip corresponding to each scene.
    \item \emph{Frame Extractor} to sample frames from each scene.
    \item \emph{Speech-to-Text} for audio transcription.
    \item \emph{Object Detector} to annotate frames with objects of interest that are related to the query.
    \item \emph{Multi-modal LLM} (or LMM) to answer the user query given the processed frames and audio transcript.
\end{enumerate}

The resulting agentic workflow for video question/answering (Q/A) can be modeled as a directed acyclic graph (DAG), where nodes represent individual agents (\eg{} Speech-to-Text transcription agent) and edges capture the data flow and execution dependencies among the agents (\eg{} passing frames from Frame Extractor to Object Detector).

\myparagraph{Code Generation}
An important use case for agentic workflows is coding and software engineering~\cite{anthropicClaudeCode,cursorFeatures2025,githubCopilotFeatures,nam2024codeunderstanding} or generating code on-the-fly for accomplishing tasks in a workflow~\cite{wang2024executable}.
We present a representative text-only workflow that translates natural language descriptions into executable Python code (\Cref{fig:code-gen-workflow}).
It adopts the \emph{LLM Debate} framework~\cite{liu2024dynamic}, where coder agents propose candidate solutions and tester agents generate tests and execute them using a Python interpreter.
Each agent plays a unique role (\eg{} algorithm developer, unit tester) and may employ the same or different LLMs.
The agents engage in iterative rounds of debate, aiming to reach consensus or terminating after a predefined number of rounds.
The final output is selected as the highest-voted solution, determined by an LLM based on both the proposed candidates and the original query.

\begin{figure}[t]
    \centering
        \begin{minipage}{\columnwidth}
            \begin{lstlisting}[style=py-acm,
            caption={Simplified video Q\slash A workflow definition. It shows tightly coupled application logic and execution details typical of existing frameworks.},
            label={lst:video-qa-workflow-definition-today}]
# ===== Workflow nodes (coupled app logic + configs) =====
scene_detection = Tool(
    fn=SceneDetector(),    # Implemented earlier by the developer.
    key=AWS_KEY, resources={"CPUs": 32}
)
frame_extractor = Tool(
    fn=FrameExtractor(),   # Implemented earlier by the developer.
    key=AWS_KEY, params={"num_frames": 15}, resources={"CPUs": 32}
)
speech_to_text = MLModel(
    name="Whisper", key=OPENAI_API_KEY, resources={"PTUs": 50}
)
object_detection = MLModel(
    name="CLIP", key=AZURE_KEY, resources={"CPUs": 128}
)
question_answer = LLM(
    name="Llama-3.2", key=DATABRICKS_API_KEY,
    params={"batch": 256}, resources={"GPUs": 8, "Type": "H100"},
    system_prompt="You are an agent that can understand videos.",
    user_prompt="Answer the given question about the video."
)
# ===== Query and data =====
ques    = "What is the name of the person wearing the red dress?"
videos  = ["road_trip.mp4"]
# ===== Workflow (ordering of nodes/data flow) =====
scenes, audio  = scene_detection(videos)
frames         = frame_extractor(scenes)
transcript     = speech_to_text(audio)
obj_frames     = object_detection(frames)
answer         = question_answer(ques, transcript, obj_frames)
            \end{lstlisting}
    \end{minipage}
    \vspace{-10pt}
\end{figure}

\subsection{Rigid and Imperative Definitions}
\label{sec:bg:workflow-def}

Current state-of-the-practice for agentic workflow development are frameworks such as LangGraph~\cite{langgraph}, LangChain~\cite{langchain}, LlamaIndex~\cite{llamaindex}, AutoGen~\cite{autogen} and CrewAI~\cite{crewai2025}.
They follow an \textit{\textbf{imperative}} paradigm:
developers define workflow components (\ie{} agents), their execution logic, interactions, and numerous \textit{\textbf{configuration parameters}}.

For example, in a video Q/A workflow (\Cref{lst:video-qa-workflow-definition-today}), the developer specifies parameters for \texttt{FrameExtractor} (number of frames per scene), enables options like speech-to-text or object detection, and defines the \textit{\textbf{execution mechanism}}, or model, for each component (\eg{} \texttt{SceneDetector} in OpenCV, Whisper~\cite{whisper-large-v3} for transcription, Llama-3.2~\cite{llama3} for reasoning).

Developers must also make \textit{\textbf{resource allocation}} decisions, such as assigning CPUs/GPUs for self-hosted or rented VMs, or choosing service tiers (\eg{} provisioned throughput units or PTU~\cite{azure-ptu}) for managed services.

This tightly couples configuration choices with the workflow DAG, making deployed workflows \emph{rigid}.
Any change requires manual updates and redeployment across the stack.

\begin{figure*}[t]
    \centering
    \begin{subfigure}[b]{0.24\textwidth}
        \includegraphics[width=\textwidth]{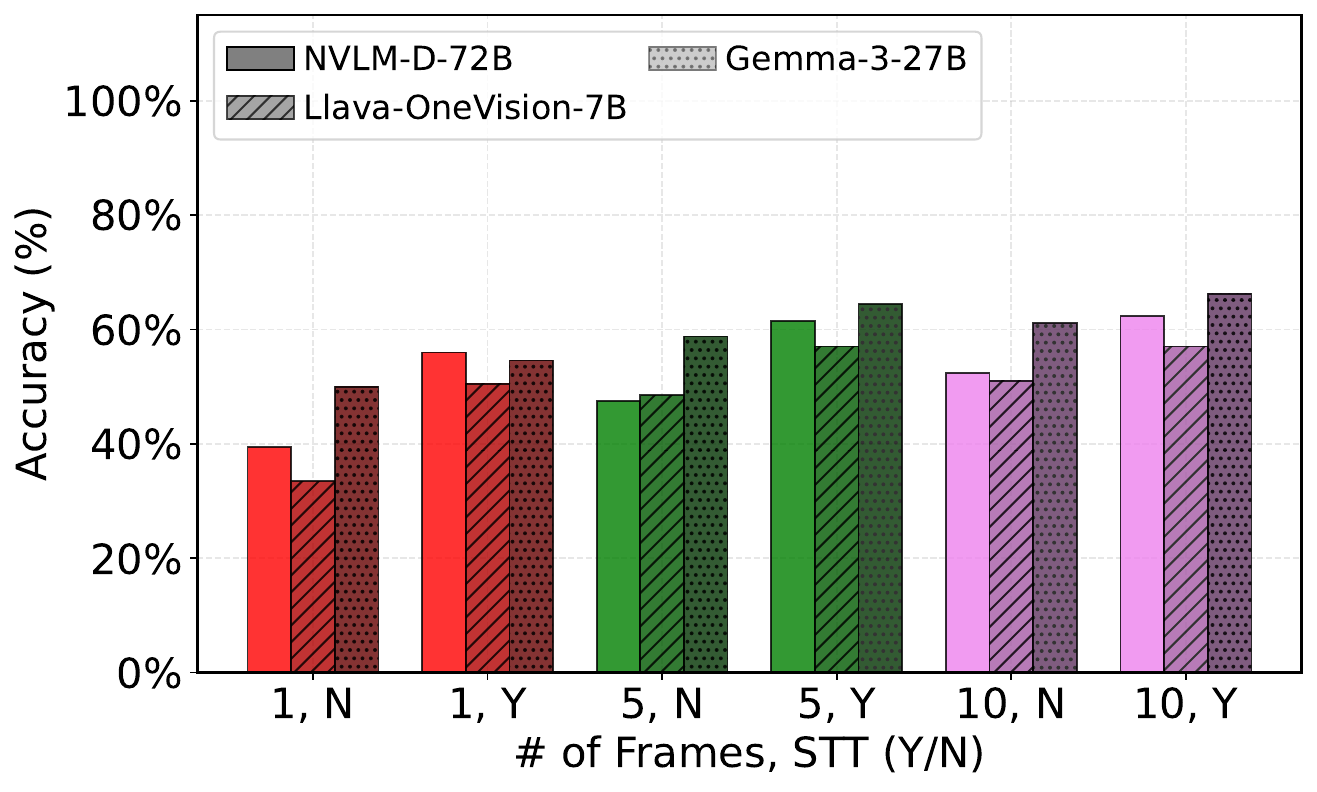}
        \caption{Video Q/A accuracy.  Varying frames, enable STT, and LMM.}
        \vspace{-8pt}
        \label{fig:video-accuracy}
    \end{subfigure}\hfill
    \begin{subfigure}[b]{0.24\textwidth}
        \includegraphics[width=\textwidth]{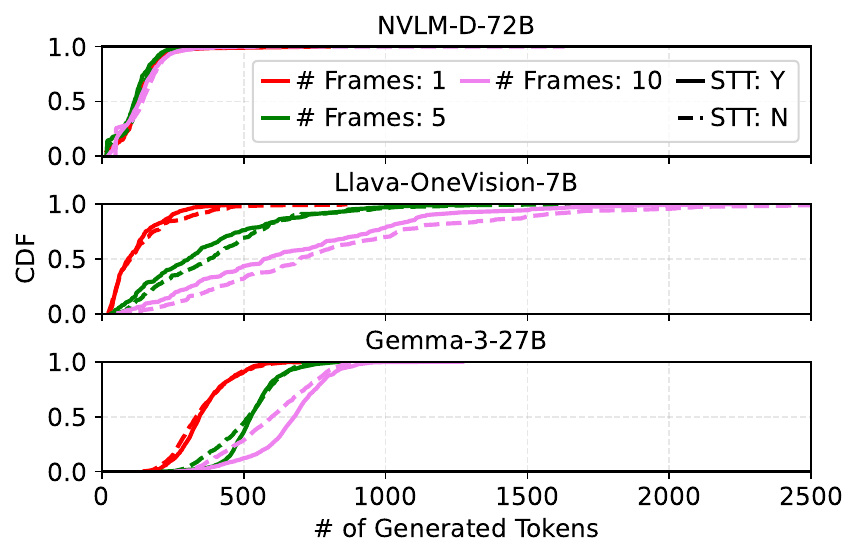}
        \caption{Video Q/A token generation by varying the knobs.}
        \vspace{-8pt}
        \label{fig:video-num-completion-tokens}
    \end{subfigure}\hfill
    \begin{subfigure}[b]{0.24\textwidth}
        \includegraphics[width=\textwidth]{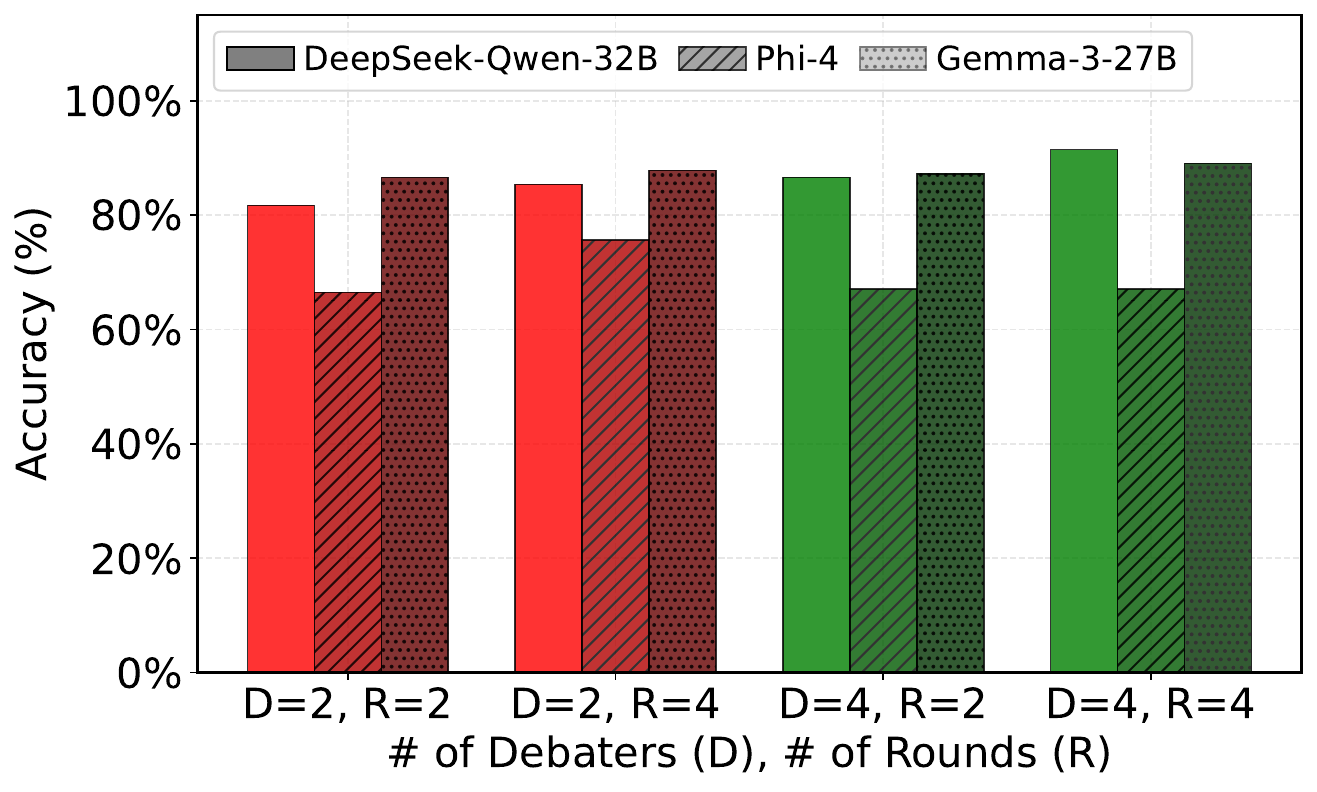}
        \caption{Code Gen. accuracy. Varying debaters, rounds, and LLM.}
        \vspace{-8pt}
        \label{fig:code-generation-accuracy}
    \end{subfigure}\hfill 
    \begin{subfigure}[b]{0.24\textwidth}
        \includegraphics[width=\textwidth]{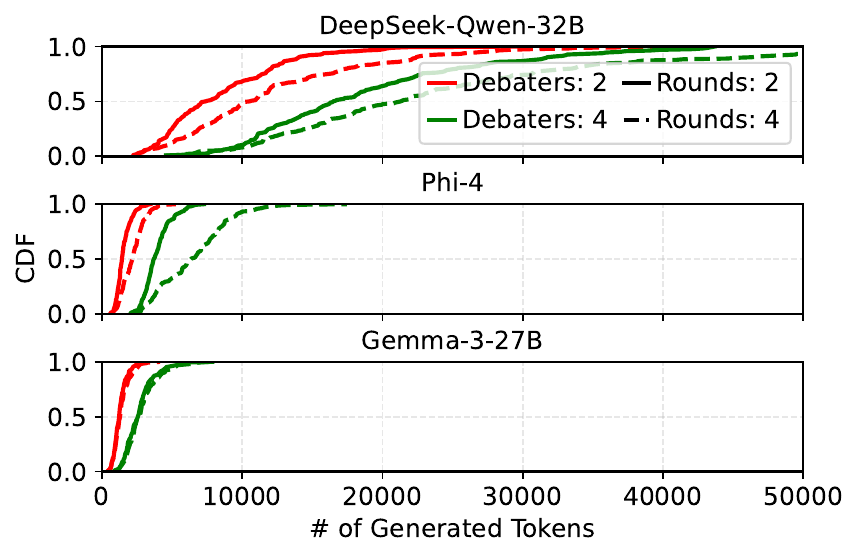}
        \caption{Code Gen. tokens generated by varying the knobs.}
        \vspace{-8pt}
        \label{fig:code-generation-num-completion-tokens}
    \end{subfigure}
    \caption{Workflow accuracy under different configurations and the token generation load on the respective models.   }
    \label{fig:workflow-profile}
\end{figure*}

\begin{figure}
    \includegraphics[width=0.49\textwidth]{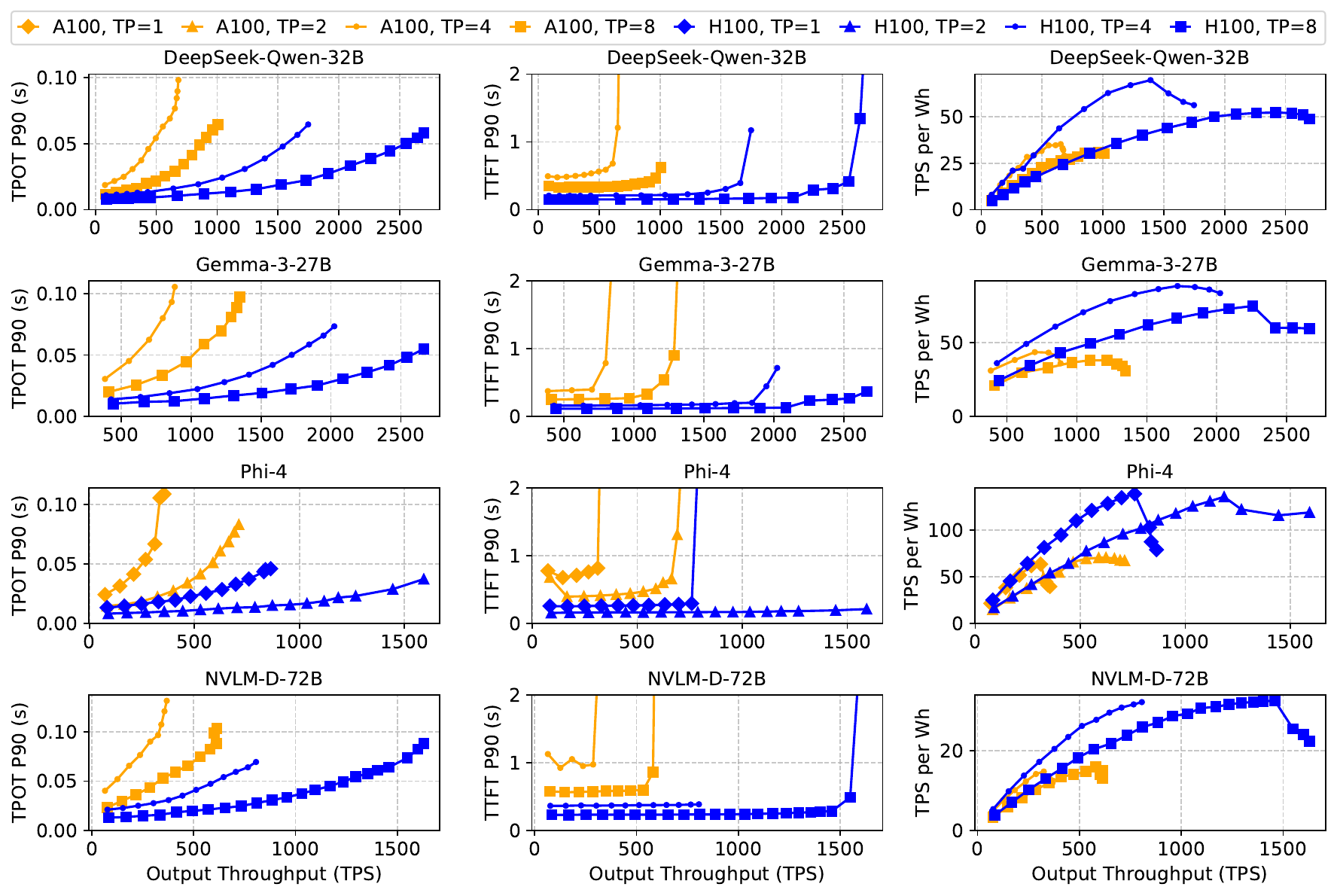}
    \caption{Model performance under different hardware and parallelism configurations.}
    \label{fig:model-profiles}
\end{figure}

\subsection{Fragmented Stack and Goals}
Agent developers develop workflows that expose endpoints callable by end users.
The workflows run on hardware offerings from cloud providers.
Entities involved in developing, using, and executing agentic workflow (\ie{} developers, end users, and providers) prioritize different goals.
\emph{Developers} value response quality, latency, or both.
\emph{End-users} additionally care about execution cost.
\emph{Providers} prioritize resource efficiency and cost.
However, the decision-making burden for workflow configuration today largely falls on developers, who must navigate an overwhelming design space and deployment knobs without holistic system visibility.
Ultimately, these decisions impact all involved entities in the agentic workflow life-cycle.
These configurations include:
\begin{itemize}[leftmargin=*]
    \item \textit{\textbf{Workflow-level configurations}}: \eg{} whether to include a Speech-to-Text Transcript agent.
    \item \textbf{\textit{Agent/Node-level configurations}}: \eg{} how many frames to extract in Frame Extractor, which LLM to use for Q/A.
    \item \textbf{\textit{Hardware-level configurations}}: \eg{} CPU \vs{} GPU and parallelism degree for each model.
\end{itemize}

\subsection{Characterization and Motivation}

\myparagraph{Poor Resource Efficiency}
Frameworks for agentic workflows (\eg{} LangGraph~\cite{langgraph} and LlamaIndex~\cite{llamaindex}) place the burden of workflow configuration on developers.
Most developers are neither systems nor ML experts, and cannot reason about accelerator choice, model parallelism, or model selection.
Configurations are therefore often arbitrary and inefficient, leading to poor performance and high cost. Even experienced developers, lacking insight into user priorities, default to maximizing accuracy at the expense of efficiency.

From the cloud provider's perspective, these workflows are opaque.
The cloud platform has little visibility into workflow components (\eg{} models) or interactions (\eg{} task sequences and data flow).
Tightly coupled application logic and execution details prevent the platform from reconfiguring workflows to improve resource efficiency while meeting SLOs on accuracy or latency.
The result is resource underutilization, increased costs, and a larger energy footprint (often passed on to users through higher prices).

\boxinsight{Cloud platforms lack visibility into workflow internals (\eg{} tasks, requirements), preventing end-to-end optimization.
Meanwhile, developers lack control or insight into system-level resource behavior.}

\myparagraph{Multi-level Trade-offs}
Configurations spanning across workflow level, agent level, and hardware layer introduce complex trade-offs over accuracy, latency, energy, and cost.

\myparagraphemph{Workflow- and Agent-Level knobs}
\Cref{fig:workflow-profile} shows a subset of configurations for the video Q/A and code generation workflows and their response accuracy.
For video Q/A, \Cref{fig:video-accuracy} shows that both workflow-level knobs (\eg{} whether to enable speech-to-text) and agent-level knobs (\eg{} number of frames extracted) significantly affect accuracy.
\Cref{fig:video-num-completion-tokens} shows the tokens generated for each configuration.
Higher token generation is associated with higher latency, cost, and GPU load.
For example, using Gemma-3-27B~\cite{gemmateam2025gemma3technicalreport} with 10 extracted frames and STT enabled achieves the highest video Q/A accuracy (66.2\%) but also generates the most tokens compared to other configurations of the same model.
Token counts vary widely across requests, from 250 to nearly 1000, often with a heavy tail.

For the code generation workflow, accuracy is sensitive to the number of debaters and rounds in LLM Debate (\Cref{fig:code-generation-accuracy}).
Some configurations, such as DeepSeek-Qwen-32B~\cite{deepseekai2025deepseekr1incentivizingreasoningcapability} (a reasoning model), achieve the highest accuracy but at a much higher token generation cost (\Cref{fig:code-generation-num-completion-tokens}).
At the median, it generates $\approx$20,000 tokens versus $\approx$2,500 for Gemma-3-27B under the same workflow configuration.
Yet token counts alone do not capture the full end-to-end workflow characteristics, such as latency, cost, or energy.

\begin{figure*}[t]
    \centering
    \begin{subfigure}[b]{0.49\textwidth}
        \includegraphics[width=\textwidth]{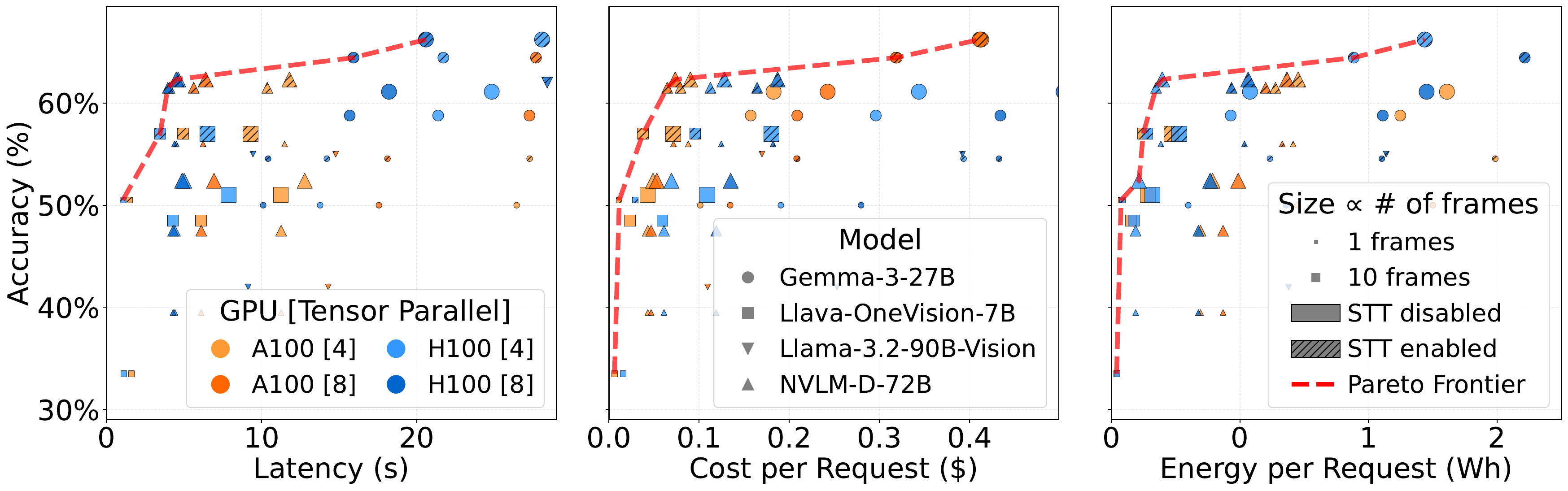}
        \caption{Video Q/A workflow.}
        \vspace{-8pt}
        \label{fig:video-pareto}
    \end{subfigure}\hfill
    \begin{subfigure}[b]{0.49\textwidth}
        \includegraphics[width=\textwidth]{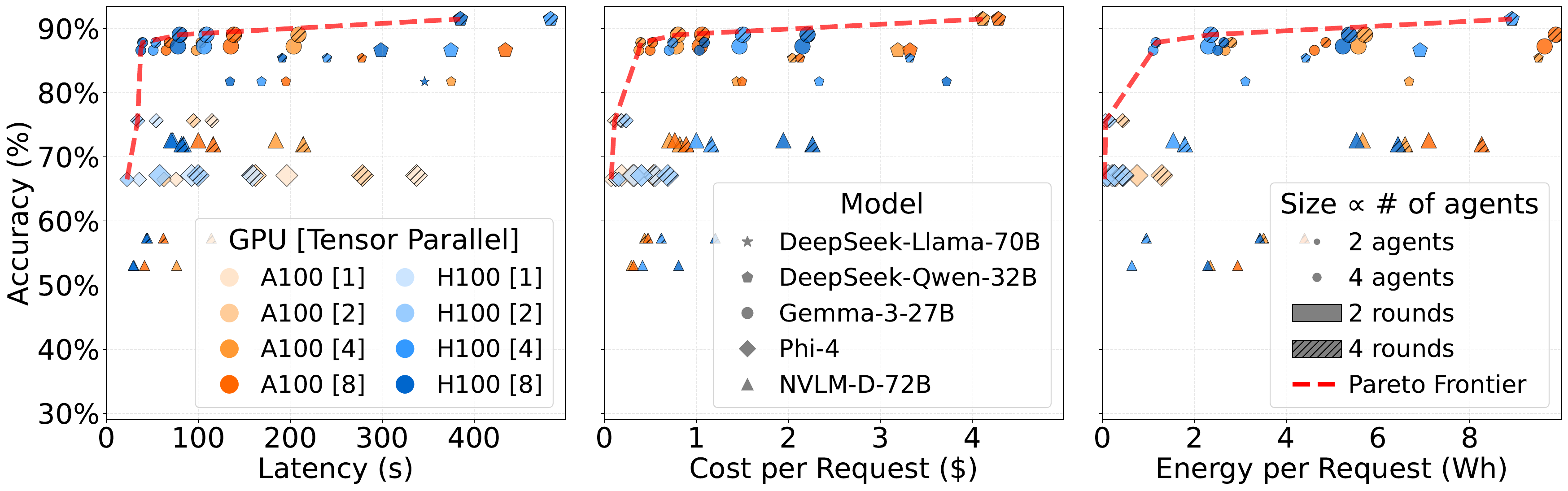}
        \caption{Code generation workflow.}
        \vspace{-8pt}
        \label{fig:code-gen-pareto}
    \end{subfigure}\hfill
    \caption{Large space of workflow configurations along a subset of knobs and metrics.}
    \label{fig:workflow-config-space}
\end{figure*}

\textit{Hardware-level knobs.}
Different models in a workflow have varying compute/memory requirements.
Their interaction with the hardware determines whether a workflow can meet user SLOs while satisfying cloud platform constraints, like the number of GPUs allocated.
\Cref{fig:model-profiles} shows models running on different hardware configurations (\eg{} GPU types and parallelism).
We evaluate time to first token (TTFT) and time per output token (TPOT) under varying load, and measure energy efficiency as throughput per Wh.
The results reveal a rich trade-off space, where configurations offer different balances of throughput, latency, and energy efficiency.

\boxinsight{Configuration knobs introduce fundamental trade-offs: workflow- and agent-level knobs affect accuracy \vs{} latency/cost, while additionally, hardware-level knobs drive cost \vs{} performance trade-offs, making it challenging to optimize workflows across multiple objectives.}

\myparagraph{High-Dimensional Configuration Space}
The configuration space for the end-to-end agentic workflow can grow exponentially with the configuration choices of the individual agents that make up the workflow.
\Cref{fig:video-pareto,fig:code-gen-pareto} show that even modest agentic workflows like video Q/A or code generation, with only a small set of knobs, produce explosive configuration spaces to choose from.
Each point in these Pareto frontiers represents a valid end-to-end configuration derived from varying model selection, hardware accelerator assignments, and resource scaling (with model parallelism) strategies.
Navigating this configuration space manually is infeasible given that:
(1) objectives vary across users (\eg{} low-latency \vs{} low-cost \vs{} high accuracy),
(2) the optimal configuration depends on dynamic runtime context (\eg{} user traffic, model availability, cloud resource availability), and
(3) model and hardware choices are rapidly evolving.

\boxinsight{Configuration complexity of each workflow grows combinatorially with the number of agents and their parameters at each level: ($O(\#WorkflowKnobs \times \#AgentKnobs \times \#HardwareKnobs)$).}
\section{\sysname{} Design}\label{sec:design}

In light of this characterization, we propose \sysname{}, a system that:
(1) has visibility into workflow requirements,
(2) holistically controls the entire workflow development and execution stack, and
(3) enables automated configuration.

Instead of rigid, imperative workflows hand-tuned by developers, \sysname{} supports \emph{declarative} specifications that \emph{decouple} high-level intent from low-level configuration.
Developers focus on application logic, while the system \emph{dynamically configures} workflows to adapt to evolving runtime conditions.
\emph{Developers} define workflow structure and task dependencies, \emph{end-users} specify SLOs (\eg{} quality, latency, cost), and the \emph{platform} dynamically optimizes execution for resource efficiency (\eg{} energy, cost) in a multi-tenant environment.
Thus, workflows can automatically adapt to new models, varying load, or changing resources (without code rewrites or redeployments).

\begin{figure*}[t]
  \centering
  \begin{subfigure}[b]{0.32\linewidth}
    \includegraphics[width=\textwidth]{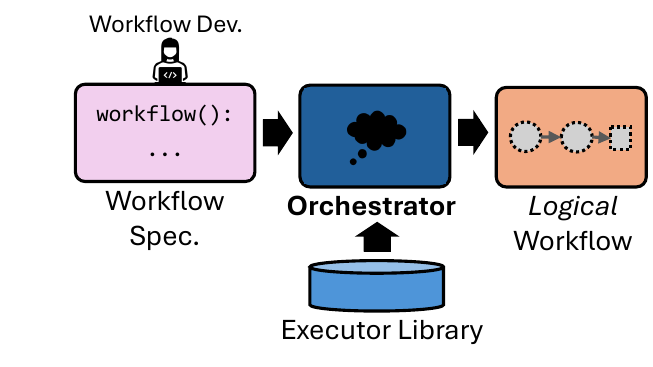}
    \caption{Development phase.}
    \vspace{-8pt}
    \label{fig:murakkab-design-deployment}
  \end{subfigure}\hfill
  \begin{subfigure}[b]{0.32\linewidth}
    \includegraphics[width=\textwidth]{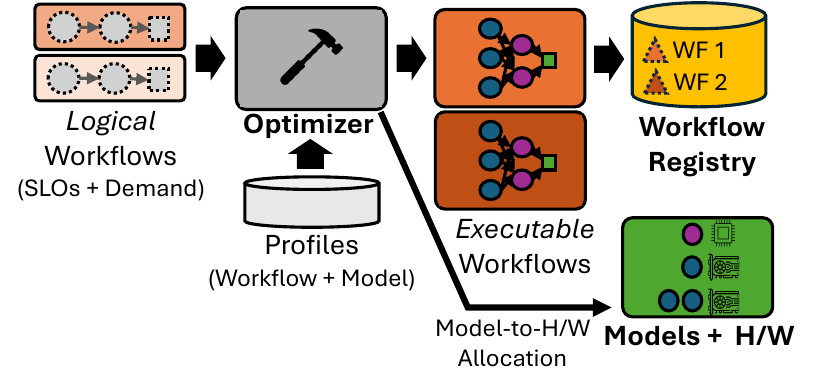}
    \caption{Optimization and deployment phase.}
    \vspace{-8pt}
    \label{fig:murakkab-design-opt}
  \end{subfigure}\hfill
  \begin{subfigure}[b]{0.32\linewidth}
    \includegraphics[width=\textwidth]{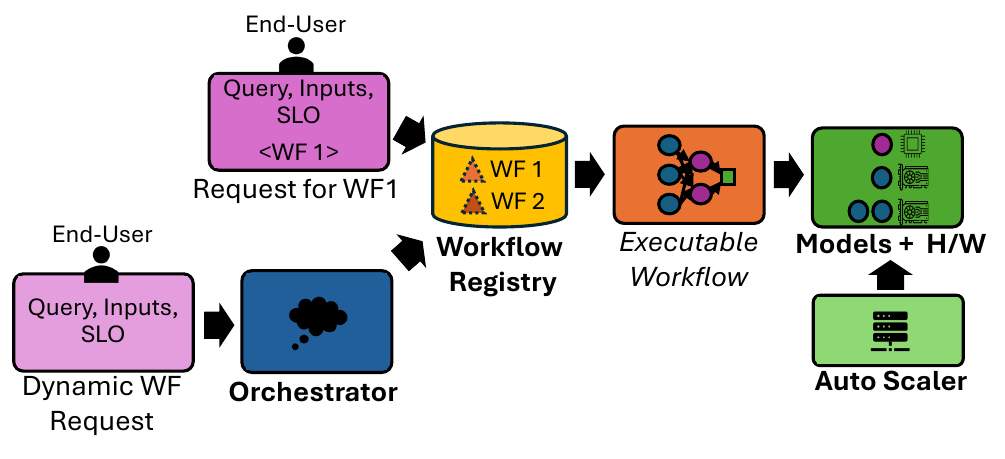}
    \caption{Execution phase.}
    \vspace{-8pt}
    \label{fig:murakkab-design-exec}
  \end{subfigure}
  \caption{\sysname{} manages end-to-end workflow life-cycle: from development to optimized deployment and execution.}
  \label{fig:murakkab-design}
\end{figure*}

\subsection{Workflow Life-Cycle}
\sysname{} cohesively manages the end-to-end life-cycle, unlike existing systems that fragment these among different entities.
There are three main phases in an agentic workflow's life-cycle, as shown in \Cref{fig:murakkab-design}: (1) Workflow Development (\Cref{fig:murakkab-design-deployment}), (2) Workflow Optimization (\Cref{fig:murakkab-design-opt}), and (3) Workflow Execution (\Cref{fig:murakkab-design-exec}).

\subsection{Development}
\label{sec:design:workflow-specification}

\sysname{} adopts a declarative paradigm for workflow specification that decouples \emph{application logic} from low-level \textit{execution details}.
Workflow developers may specify high-level tasks, and optionally, the data flow between them, without the need to provide any workflow configurations (\eg{} resource allocation or model selection details).

\begin{figure}[t]
    \centering
        \begin{minipage}{\columnwidth}
            \begin{lstlisting}[style=py-acm,
            caption={\sysname{}'s declarative workflow specification of the video Q/A abstracts away configuration details, letting developers focus on application logic.},
            label={lst:murakkab-static-workflow}]
# ===== Sub-tasks in the workflow =====
scene_detect   = "Given a list of videos, identify scenes in each."
frame_extract  = "Given a list of scenes, extract frames."
stt            = "Given a list of scenes, convert audio to text."
q_a            = "Answer the query given some context."
# ===== Workflow description using the sub-tasks and data flow =====
def workflow(query, videos):
    scenes      = scene_detect(videos)
    frames      = frame_extract(scenes)
    transcript  = stt(scenes)
    answer      = q_a(query, [frames, transcript])
    return answer
# ===== Execution with example request =====
query   = "What is the name of the person wearing the red dress?"
videos  = ["road_trip.mp4"]
result  = run(workflow(query, videos), slo=LOW_LATENCY)
            \end{lstlisting}
    \end{minipage}
    \vspace{-8pt}
\end{figure}

\myparagraph{Executor Library}
\sysname{} is designed to interoperate with existing model and tool ecosystems.
It supports LLMs and traditional machine learning (ML) models from repositories such as Hugging Face~\cite{huggingface}, as well as tools from open-source libraries and platforms, including OpenAI Agents SDK~\cite{openai-sdk}, Google Vertex AI Agent Garden~\cite{agent-garden}, NVIDIA NeMo Agent Toolkits~\cite{nemo-sdk}, and Microsoft Azure AI Foundry Tools~\cite{azure-ai-foundry-tools}.
In \sysname{}, an \emph{executor} is a functional unit within a workflow that can take one of three forms:

(1) \textbf{\textit{LLM}}: specialized LLM configurations (fine-tuning, few-shot learning, or even just domain-specific prompting);

(2) \textbf{\textit{Structured compositions}}: aggregations of models, \eg{} a self-reflection~\cite{renze2024self} or an LLM-Debate pattern~\cite{liu2024dynamic} built from multiple LLMs; 

(3) \textbf{\textit{Tool}}: utility modules for AI workflows to take actions with (\eg{} OpenCV frame extractor for video processing, web-search, file-search, computer-use, or any third-party tools that follow the MCP specification~\cite{mcp}).
Traditional ML models are also included as tools (\eg{} CNN-based image classifiers or Word2Vec sentiment analyzers).

\myparagraphemph{Attributes}
Each model or tool in the library exposes three attributes:
(1) a textual description,
(2) an interface specification, and
(3) a key-value list of configurable parameters.
For example, the frame extraction tool exposes the knobs:
\texttt{F} (number of frames to extract) and \texttt{cores} (number of CPU cores to run on).
The LLM Debate composition exposes the knobs:
\texttt{D} (number of debaters), \texttt{R} (number of rounds), and \texttt{model} (which LLM to use).
The \emph{orchestrator} uses these descriptions and interfaces to rank and assign executors (models or tools) for workflow tasks, deferring parameter configuration to a later optimization phase (\Cref{sec:design:optimization}).

A key design choice in \sysname{} is mapping a broad range of unknown tasks to executors that are built from a finite, known set of models and tools in the library.
If none is found, \sysname{} prompts the developer to onboard a suitable one.

\myparagraph{Workflow Specification}
We allow tasks to be expressed in \emph{natural language} with varying levels of control over how the task may be executed, in contrast to imperative paradigms discussed earlier.
Recent agentic-workflow development frameworks (\eg{} DSPy~\cite{dspy}) are moving in this direction.
\sysname{} takes this a step further and completely decouples configuration details from the workflow specification.

\myparagraphemph{Declarative Specification}
Developers can manually decompose tasks into sub-tasks and define data flow between them, providing precise control over workflow steps.
\Cref{lst:murakkab-static-workflow} shows the developer specifies sub-tasks (\eg{} scene detection, speech-to-text transcription, frame extraction) and their dependencies, ensuring sequential execution to achieve the desired outcome, such as answering a question about a video.
Configuration details (\eg{} which LLM to use, number of frames to extract, resource allocation) are omitted from the specification.
However, \sysname{} does not restrict developers from specifying any execution preferences (\eg{} particular LLM choice or hardware constraint), which are then incorporated into the optimization process as constraints.

\myparagraphemph{Interface}
Once deployed, workflows are exposed to end users via dedicated REST endpoints.
Each request may include an SLO, such as accuracy, latency, or cost tier.
\sysname{} uses these SLOs to configure workflows per request and optimize resource efficiency across tasks.

\myparagraph{Workflow Orchestrator}
The \emph{workflow orchestrator} transforms a \emph{declarative workflow} specification into a \emph{logical workflow}.
It interprets the specification, parses tasks and sub-tasks, and maps each to an appropriate executor from \sysname{}'s library.
At the core is an LLM with tool-calling capabilities~\cite{qin2023toolllm}, which receives a list of available executors and their interfaces, along with task descriptions, and selects the best executor for each sub-task.
Recent work~\cite{niu2025flow,li2024autoflow,zhang2024aflow} has focused on automating the discovery and refinement of effective agentic workflows to maximize task performance and minimize human intervention, which \sysname{} can benefit from.
A feedback loop allows developers to inspect and refine the generated specification, supporting hybrid workflows with both manual and system-generated tasks.

\myparagraph{Logical Workflow}
This is an abstract execution plan that captures the functional intent of each task without binding to specific models, resources, or hardware.
It is represented as a directed acyclic graph (DAG), where nodes correspond to executors and edges indicate data flow.
This representation remains request-agnostic, containing no per-request details such as query text, input payloads, or SLOs.
Execution specifics (\eg{} model selection or hardware allocation) are deferred to later stages.

The orchestrator performs type-checking on the DAG to ensure output types from source nodes match input types of destination nodes.
In case of mismatches, the workflow is regenerated with error feedback to the LLM.
Persistent errors prompt the developer to revise the specification.

\subsection{Optimization and Deployment}
\label{sec:design:optimization}

\myparagraph{Profiles}
Accurate, fine-grained performance characterization is essential for optimizing multi-tenant agentic workflows with dynamic execution patterns.
Inspired by Profile-Guided Optimization (PGO)~\cite{wintermeyer2020p2go,samplepgo}, \sysname{} builds offline profiles across diverse configurations to inform runtime decisions.
Each profile captures three key metrics per workflow configuration:
response quality, end-to-end latency, and resource usage.
To support this, \sysname{} maintains two profiling layers:
workflow profiles and model profiles.

\myparagraphemph{Workflow Profiles}
\sysname{} profiles representative agentic workflows (\eg{} video Q/A) under various combinations of workflow-level knobs (\eg{} enabling/disabling STT) and executor-level knobs (\eg{} varying the number of frames extracted by the extractor tool).
\emph{Quality} is assessed using open-source benchmarks and datasets (\eg{} VideoMME~\cite{fu2025videommefirstevercomprehensiveevaluation}, HumanEval~\cite{chen2021evaluatinglargelanguagemodels}, and Math~\cite{hendrycksmath2021}) with ground-truth results.
When public datasets fall short, \sysname{} can onboard curated evaluation datasets from developers or collect request-response pairs and user feedback (\eg{} rating, approvals), similar to deployed systems like ChatGPT~\cite{openai2023chatgpt}.
These inputs are periodically incorporated into profile updates to refine configuration choices for active workflows.

Configuration choices also affect per-request resource demands.
Workflow profiles quantify executor-level load, including prompt and completion tokens for LLM-based executors, serving as a proxy for resource usage.
These profiles capture metrics similar to those in our earlier characterization (\eg{} \Cref{fig:video-accuracy,fig:video-num-completion-tokens}).

\myparagraphemph{Model Profiles}
To assess workflow resource demands, executor-level load must be mapped to model and hardware metrics.
\sysname{} maintains profiles for each model, capturing performance across software configurations (\eg{} tensor parallelism, prefill/decode separation) and hardware setups (\eg{} GPU type, clock frequency).
Each profile reports:
(1) latency (TTFT and TPOT for LLMs),
(2) energy consumption across hardware, and
(3) cost per configuration.
Profiles span load levels to expose trade-offs and guide the optimizer in allocating load and instances.

Profiling is lightweight; performed once per configuration and reused across workflows.
New models and accelerators are profiled upon integration.
Decoupling workflow and model profiles enables workflows to benefit immediately from model updates, with selective re-profiling as needed.

This separation enables independent evolution, profile reuse, rapid model onboarding, and efficient hardware adaptation.
The optimizer uses profiles as structured priors to evaluate cost--accuracy--latency trade-offs and drive reconfiguration under dynamic conditions.

\myparagraph{Workflow Optimizer}
The optimizer transforms the \emph{logical workflow} from the orchestrator into an \emph{executable workflow} by selecting:
(1) workflow parameters,
(2) models/tools per executor, and
(3) hardware and parallelism strategies.

It combines workflow and model profiles to estimate latency, cost, and energy for candidate configurations.
Workflow profiles capture task accuracy and executor-level load; model profiles quantify latency, energy, and cost under varying load and system setups.
For requests with SLOs on accuracy, latency, or cost, the optimizer selects configurations that meet constraints while maximizing efficiency.

\begin{figure}
    \includegraphics[width=\columnwidth]{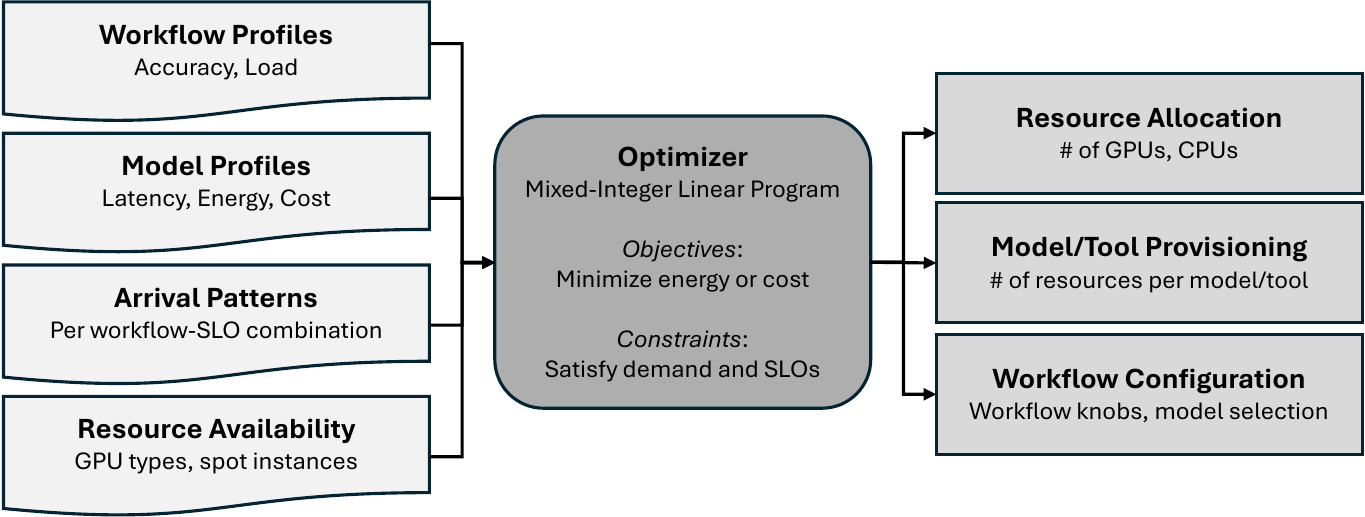}
    \caption{Summary of \sysname{}'s optimization process.}
    \label{fig:optimizer-summary}
    \vspace{-4pt}
\end{figure}

For example, to estimate latency for a video Q/A request, \sysname{}:
(1) filters out configurations that do not meet the quality SLO, % using workflow profiles,
(2) uses token distributions and model profiles to estimate latency and cost,
(3) selects a configuration satisfying all SLOs (or returns an infeasibility error).

Beyond individual workflows, \sysname{} optimizes system-wide resource usage.
It leverages global visibility to align configurations across workflows, enabling colocation and executor multiplexing for diverse SLOs.

During each \emph{optimization epoch}, the optimizer considers workflow load and resource availability (\eg{} spot instances) to allocate model instances.
It balances peak and average load to maximize sharing while respecting SLOs.
\Cref{fig:optimizer-summary} summarizes the optimization process.

\myparagraphemph{Formulation}
The optimizer uses a Mixed Integer Linear Programming (MILP) formulation to allocate resources for agentic workflows under varying SLOs.
Inputs include:
(1) \emph{workflow profiles} defining accuracy, token needs, and SLO targets (latency, cost, accuracy),
(2) \emph{model profiles} specifying throughput, latency, energy, GPU requirements,
(3) \emph{arrival patterns} showing request distributions per workflow-SLO pair, and
(4) \emph{resource constraints} including cost limits.

The MILP matches workflows to feasible model profiles, then determines instance counts and workload distribution.
A key insight is separating peak provisioning from average utilization:
GPUs are allocated for peak demand, while average loads guide cost-efficient usage.

The optimizer supports multiple objectives: minimizing energy, minimizing cost, or maximizing accuracy given a cost budget.
The output is a deployment plan:
GPU instance counts per model profile and workflow-to-instance assignments, ensuring SLO compliance and resource efficiency.
We provide the detailed formulation in \Cref{sec:optimization-milp}.

Workflows that can benefit from \textit{external} calls to proprietary models can also be supported by the optimizer by assigning appropriate cost, latency, and accuracy attributes to those model profiles.

Once an executable workflow is generated for all valid SLO tiers of an onboarded workflow, it is added to the \sysname{} workflow registry and is ready to serve requests.

\subsection{Execution}
\label{sec:design:execution}
At runtime, \sysname{} receives incoming requests from end-users with a payload that contains the identifier of the agentic workflow being invoked, any input query/data, and the SLOs.
\sysname{} looks up the registry to obtain the corresponding executable workflow and submits it for execution.

\myparagraph{Dynamic Workflow Requests}
End-users can either invoke a \emph{particular} agentic workflow or send a request with a natural language query (\Cref{fig:murakkab-design-exec}), any input data to operate on, and SLOs, \emph{without} specifying an agentic workflow to use.
The workflow orchestrator parses the query into one or more sub-tasks, mapping each to an appropriate executor or an existing workflow.
Thus, \sysname{} dynamically composes workflows from existing building blocks available to it.

\myparagraph{SLOs}
Each request has the option to specify a quality, latency and cost SLOs.
We assign four SLO tiers for quality and end-to-end latency: \emph{best}, \emph{good}, \emph{fair}, and \emph{basic}.
The SLO tiers correspond to the best, 95$^{th}$, 80$^{th}$, and 50$^{th}$ percentile values of accuracy and latency available among the set of all workflow, model, and hardware configurations.

\myparagraph{Runtime Optimization}
While request dispatch is deterministic given the selected workflow, achieving efficient execution under dynamic, multi-tenant conditions requires continuous adaptation.
The optimizer runs in the background after every optimization epoch, in our case every 60 minutes, to adapt to the most up-to-date load and resource availability in the system.
The state of the previous epochs is used to project the load for each workflow in the next epoch.
Using this information, the optimizer reconfigures the workflows and updates their executable workflows in the registry.

\myparagraph{Auto-Scaler}
Predicting end-to-end resource usage in agentic workflows is difficult, as input-dependent control flow and intermediate outputs propagating along data flows determine actual demand (\eg{} \Cref{fig:video-num-completion-tokens,fig:code-generation-num-completion-tokens}).
For example, a video Q/A workflow with 10 frames and STT on Llava-OneVision-7B~\cite{li2024llavaonevisioneasyvisualtask} produces 600 and 1200 tokens in the 50$^{th}$ and 99$^{th}$ percentile, respectively, highlighting high variance.

To handle such variability, \sysname{} includes an \emph{auto-scaler} that monitors per-model instance load over short windows (seconds to minutes) and rapidly scales out when needed.

We set thresholds for autoscaling based on the performance-throughput characteristic in executor profiles.
This mechanism prioritizes avoiding SLO violations over short-term allocation optimality.
\sysname{} also maintains spare resources to absorb demand spikes and, when it detects significant deviations in workload or resource usage, triggers early re-optimization to adapt quickly.

\section{Evaluation}\label{sec:eval}

\begin{figure}[t]
  \centering
  \begin{subfigure}[b]{0.49\linewidth}
    \includegraphics[width=\linewidth]{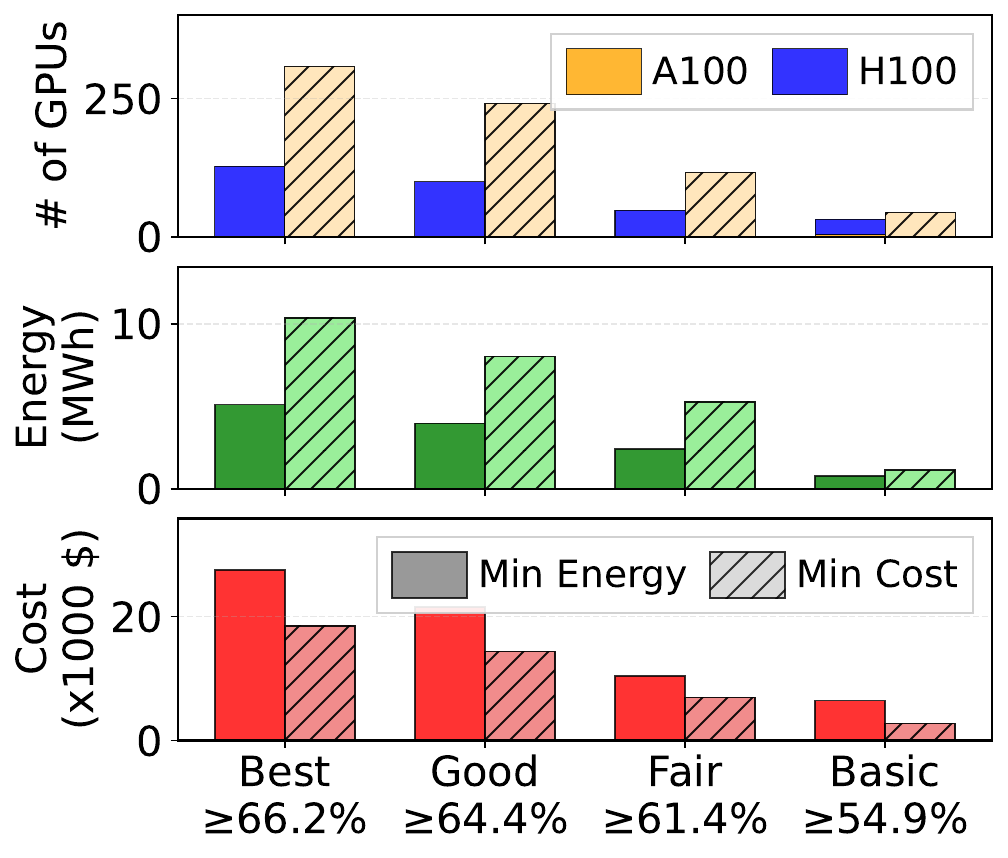}
    \caption{Accuracy SLOs.}
    \vspace{-8pt}
    \label{fig:video-accuracy-slo}
  \end{subfigure}\hfill
  \begin{subfigure}[b]{0.49\linewidth}
    \includegraphics[width=\linewidth]{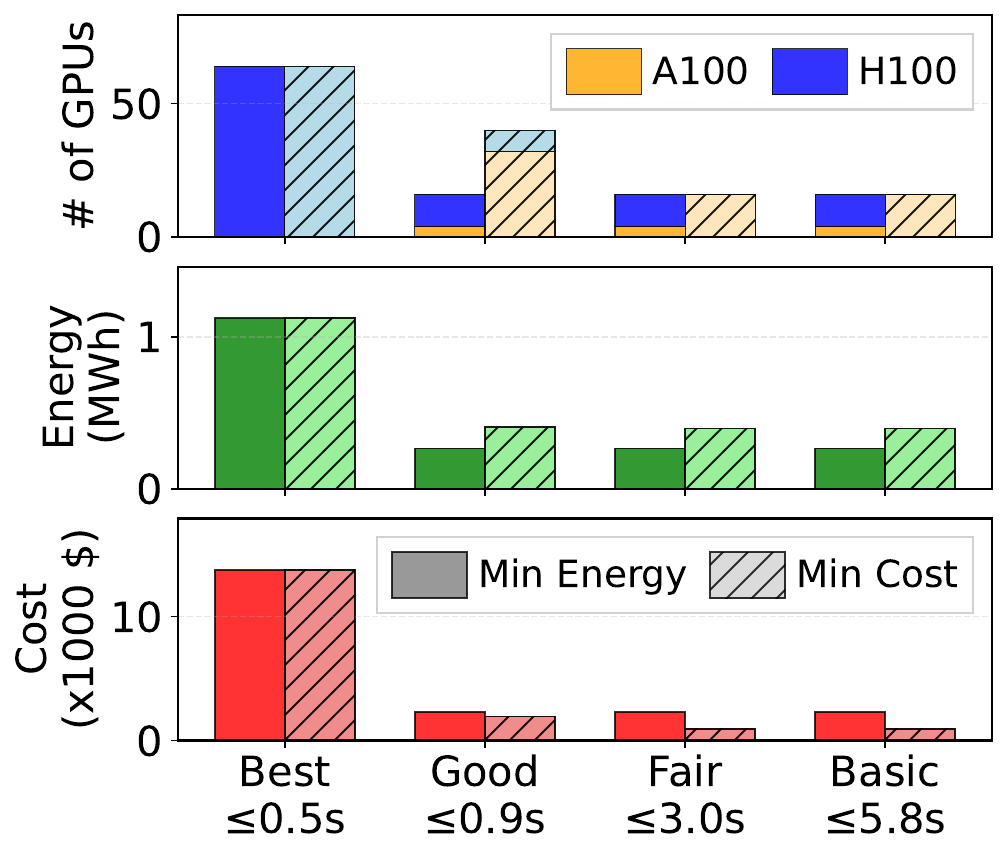}
    \caption{Latency SLOs.}
    \vspace{-8pt}
    \label{fig:video-latency-slo}
  \end{subfigure}
  \caption{Video Q/A workflow configured for different SLOs and optimization objectives.}
  \label{fig:video-optimizations}
\end{figure}

\begin{figure}
  \centering
  \begin{subfigure}[b]{0.49\linewidth}
    \includegraphics[width=\linewidth]{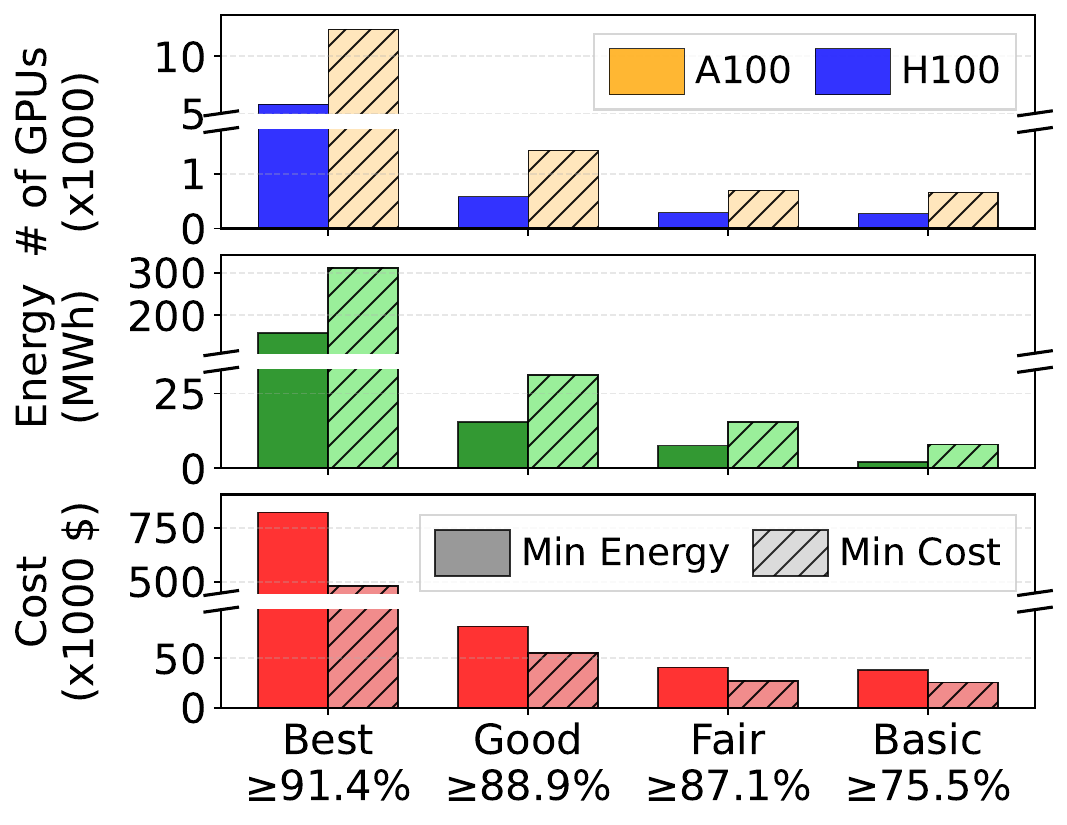}
    \caption{Accuracy SLOs.}
    \vspace{-8pt}
    \label{fig:code-gen-accuracy-slo}
  \end{subfigure}\hfill
  \begin{subfigure}[b]{0.49\linewidth}
    \includegraphics[width=\linewidth]{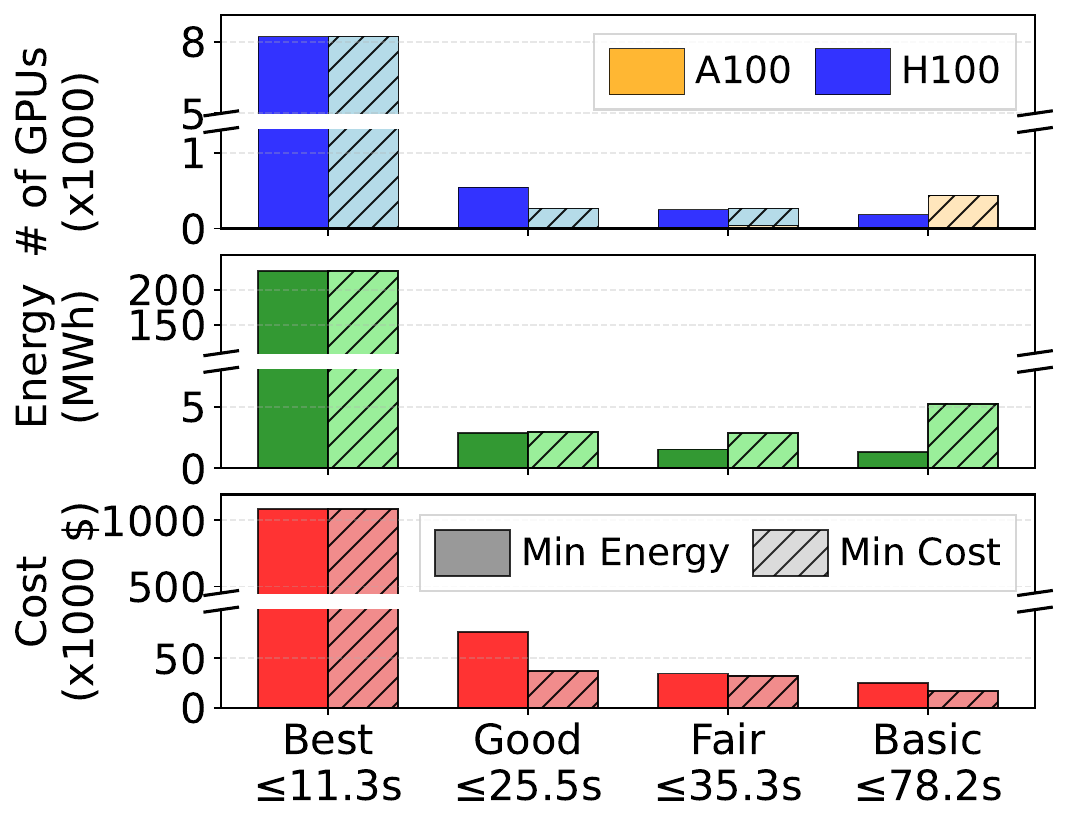}
    \caption{Latency SLOs.}
    \vspace{-8pt}
    \label{fig:code-gen-latency-slo}
  \end{subfigure}
  \caption{Code generation workflow configured for different SLOs and optimization objectives.}
  \label{fig:code-gen-optimizations}
\end{figure}

\subsection{Experimental Setup}

\myparagraph{Hardware}
We run our experiments on A100 and H100 VMs from Microsoft Azure.
Each A100 VM has 8$\times$NVIDIA A100 (80GB) GPUs and an AMD EPYC 7V12 64-Core processor, while each H100 VM has 8$\times$NVIDIA H100 (80GB) GPUs with an Intel Xeon (Sapphire Rapids) processor.
We use vLLM~(v0.9)~\cite{kwon2023efficient} as the LLM inference engine, speaches-ai~(v0.7)~\cite{speachesAI2025} as the speech-to-text model serving engine, and OmDet~\cite{zhao2024omdet} for object detection model serving.

\myparagraph{Production Traces}
Since no publicly available traces exist for production agentic workflow serving, we approximate workload arrivals using LLM serving traces collected over a 24-hour period in May 2024 from Azure’s LLM inference service for \emph{chat} and \emph{coding} applications~\cite{stojkovic2025dynamollm} (\Cref{sec:appendix:traces}).

\myparagraph{Agentic Workflows}
Our evaluation focuses on the video Q/A and code generation workflows.
We map the \emph{chat} requests from the trace to the \emph{video Q/A} workflow and \emph{coding} requests to the \emph{code generation} workflow.

\myparagraph{Policies}
We consider three scheduling policies:

(1) \emph{Static} is a hand-crafted baseline that balances cost and accuracy for both workflows, using Gemma-3-27B and A100 GPUs.
However, it lacks visibility into the workflows and has a disconnect between orchestration and resource management, resulting in lack of adaptability to shifting demand.
This represents existing systems like LangGraph~\cite{langgraph}.

(2) \emph{\sysname{} Optimized} (\sysnameshort{} Opt) optimizes each workflow--SLO combination for a specific objective (either minimizing energy or cost).

(3) \emph{\sysname{} Optimized + Multiplexing} (\sysnameshort{} Opt+Mult) \emph{jointly} optimizes requests across all workflow--SLO combinations to maximize colocation and sharing of model instances for even better efficiency.

When evaluating \sysname{} over the production traces, we set the optimization epoch to 60 minutes, and also conduct a sensitivity analysis for this choice in \Cref{sec:eval:sensitivity}.

\subsection{Single-Workflow Optimization}
\label{sec:eval:single}
To understand how \sysname{} optimizes a single workflow under varying request SLOs and optimization objectives, we run \sysname{} with two optimization objectives:
\emph{minimize energy consumption} and \emph{minimize execution cost}.
We use the workload traces~\cite{stojkovic2025dynamollm} for each workflow and assume that all requests have the same SLO for each experiment.
We report the type and number of GPUs allocated, energy consumption in \texttt{MWh}, and the cost in \texttt{\$}.

\myparagraph{Accuracy SLOs}
\Cref{fig:video-accuracy-slo} shows the requests across SLO accuracies.
When minimizing energy, \sysname{} lowers energy consumption from 5.1 MWh at 66.2\% accuracy (\emph{best}) to 3.9 MWh at 64.4\% accuracy (\emph{good}) -- an energy reduction of 23.5\% with negligible accuracy impact.
Tolerating a drop to 61.4\% (\emph{fair}) yields substantial savings of 2.6 MWh from the peak consumption.
On the other hand, when optimizing for cost, \sysname{} reduces from \$18.5k to \$14.3k while marginally dropping accuracy between \emph{best} to \emph{good}.
Allowing accuracy to drop further to 61.4\% reduces expenses to \$6.9k, a $\approx 4\times$ decrease compared to the most expensive configuration.
\sysname{} prefers using H100 GPUs when minimizing energy, trading off cost for energy savings.

\Cref{fig:code-gen-accuracy-slo} shows results for the code generation workflow.
Energy use ranges from 312 MWh to 2 MWh, and cost from \$820k to \$25k, across varying accuracy levels.
Relaxing the accuracy SLO from \emph{best} to \emph{good} yields a sharp drop in energy ($\approx 10.5\times$) and cost ($\approx 8.7\times$), mainly due to \sysname{} switching from DeepSeek-Qwen-32B to Gemma-3-27B, which has lower token and operation costs.

\myparagraph{Latency SLOs}
\sysname{} serves latency-sensitive requests with varying SLO tiers, while minimizing energy or cost.
A similar pattern in resource and cost savings emerges for both workflows as shown in \Cref{fig:video-latency-slo,fig:code-gen-latency-slo}.
We guarantee a basic accuracy tier even for latency SLO requests (\eg{} 50\% for video Q/A).
For example, \sysname{} can reduce energy consumption from 1.1 MWh to 266 kWh at a slight increase in end-to-end latency, from 0.5 s to 0.9 s for the video Q/A workflow.
On the other hand, it can reduce energy consumption from 227 MWh to 2.8 MWh (as a result of changing to a different model with different tensor parallelism).

We detail the most common workflow configurations chosen by \sysname{}'s optimizer for each of the experiments over the entire arrival trace in \Cref{sec:appendix:video,sec:appendix:code-gen} (\Cref{tab:video_configs_gpu,tab:code_generation_configs_gpu}).
We observe similar trends for the Math Q/A workflow and report the results in \Cref{sec:appendix:math} (\Cref{fig:math-accuracy-slo,fig:math-latency-slo}).

\subsection{Multi-Workflow Optimization}
\label{sec:eval:multi}

\begin{figure*}
    \centering
    \begin{subfigure}[b]{0.49\linewidth}
        \includegraphics[width=\textwidth]{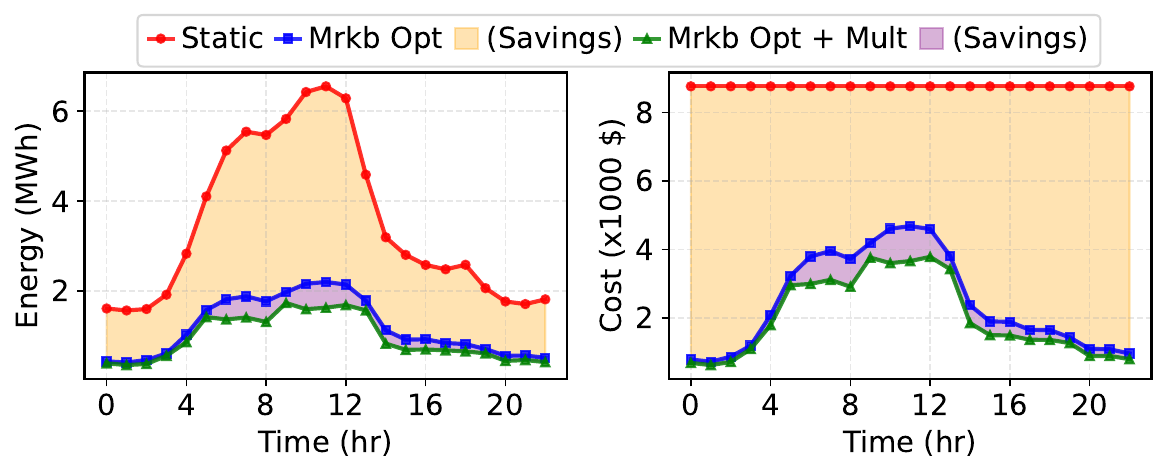}
        \caption{Energy and cost over time. Static policy is agnostic to load shifts while \sysname{}  reconfigures workflows for higher efficiency.}
        \vspace{-8pt}
        \label{fig:multiplex-static-conservative-a100-compare}
    \end{subfigure}\hfill
    \begin{subfigure}[b]{0.49\linewidth}
        \includegraphics[width=\textwidth]{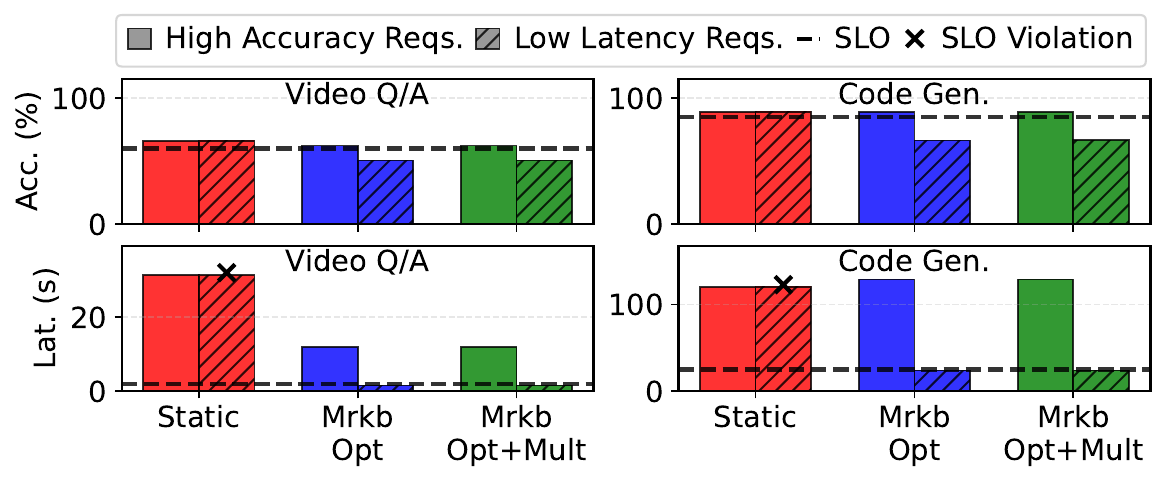}
        \caption{Accuracy and latency for all requests. Static policy does not distinguish between request SLO categories.}
        \vspace{-8pt}
        \label{fig:multiplex-static-conservative-a100-accuracy-latency}
    \end{subfigure}
    \caption{Comparing three policies: (1) a hand-crafted static configuration, (2) \sysname{} optimizing individual workflows (\emph{Opt}), and (3) \sysname{} jointly-optimizing across workflows + multiplexing resources (\emph{Opt+Mult}).}
    \label{fig:multiplex-static-conservative-a100}
\end{figure*}

We measure the improvements in efficiency as a result of joint optimization and multiplexing across different workflow--SLO combinations.
For this experiment, we run video Q/A and code generation requests together and assign 70\% requests to be high-accuracy and 30\% requests to low-latency, both with \emph{good tier} (\Cref{sec:design:execution}).

\begin{table}
\centering
\scriptsize
\setlength{\tabcolsep}{3pt}
\begin{tabular}{c c c c}
\toprule
\textbf{Policy} &
\makecell{\textbf{\# of GPUs}} &
\makecell{\textbf{Energy (MWh)}} &
\makecell{\textbf{Cost ($\times$1000 \$)}} \\
\midrule
Static        & 2560 & 80.4 & 201.5 \\
\sysname{} Opt      & 1151 & 27.1 & 56.2  \\
\sysname{} Opt+Mult & 908  & 21.6 & 46.5  \\
\bottomrule
\end{tabular}
\caption{Comparison of resource usage, energy, and cost. \sysname{} substantially reduces GPUs, energy, and cost compared to static allocation, with further gains from multiplexing across all video Q/A and code generation workflow requests.}
\vspace{-10pt}
\label{tab:multiplex-video-code-gen-static-conservative-a100-absolute}
\end{table}

\Cref{fig:multiplex-static-conservative-a100-compare} shows the \emph{static baseline} has a fixed allocation with a fixed cost.
Its energy consumption varies depending on resource usage, but is significantly higher than the other configurations due to higher idle-power consumption of its GPUs.
\Cref{tab:multiplex-video-code-gen-static-conservative-a100-absolute} summarizes the results for the 24 hour trace.
This configuration uses $\approx$ 2,560 A100 GPUs consistently resulting in a flat hourly cost.
A non-hand-crafted baseline could be \emph{significantly} more inefficient than our current baseline from requiring bigger models and more resources or result in more SLO violations if under-provisioned.

\myparagraph{\sysnameshort{} Opt}
This policy optimizes each workflow--SLO combination to minimize cost, although \emph{without} considering the demand across other combinations running in the system.
It can adapt to changing load (leveraging the difference between peak and average utilization) to change model and resource allocation over time.
It requires 1,151 GPUs, 27.1 MWh energy, and \$56,246 cost.

\myparagraph{\sysnameshort{} Opt+Mult}
This policy additionally leverages its \emph{holistic} view of all workflows and their demand to maximize multiplexing opportunities.
It requires 908 GPUs (21.1\% reduction), 21.6 MWh energy (20.2\% reduction), and \$46,494 cost (17.3\% reduction) respectively.

\myparagraph{Accuracy and Latency}
The static baseline cannot distinguish between per-request SLO variations to configure workflows appropriately.
Therefore, high-accuracy and low-latency SLO requests have the same accuracy and latency, resulting in latency SLO violations, as shown in \Cref{fig:multiplex-static-conservative-a100-accuracy-latency}.
\sysname{} leverages per-request SLO information to provision enough resources for each request, as can be seen in the difference in accuracy and latency of the two categories of requests, allowing for higher resource-efficiency.

\subsection{Adapting to Dynamic Resource Availability}\label{sec:eval:adapt}

\begin{table}[t]
\centering
\scriptsize
\setlength{\tabcolsep}{3pt}
\begin{tabular}{c c c c c c}
\toprule
\makecell{\textbf{Available}\\\textbf{A100s}} &
\makecell{\textbf{Available}\\\textbf{H100s}} &
\makecell{\textbf{Allocated}\\\textbf{A100s}} &
\makecell{\textbf{Allocated}\\\textbf{H100s}} &
\makecell{\textbf{Energy}\\\textbf{(MWh)}} &
\makecell{\textbf{Cost}\\\textbf{($\times$1000 \$)}} \\
\midrule
\cellcolor{grayHigh}2000 & \cellcolor{grayLow}0   & \cellcolor{gpuHigh}1292 & \cellcolor{hgpuLow}0   & \cellcolor{energyHigh}24.7 & \cellcolor{costLow}55.7 \\
\cellcolor{grayHigh}2000 & \cellcolor{grayLow}100 & \cellcolor{gpuHigh}780  & \cellcolor{hgpuLow}100 & \cellcolor{energyHigh}17.8 & \cellcolor{costLow}52.4 \\
\cellcolor{grayHigh}2000 & \cellcolor{grayMed}200 & \cellcolor{gpuMed}536   & \cellcolor{hgpuMed}200 & \cellcolor{energyMed}14.5  & \cellcolor{costMed}60.3 \\
\cellcolor{grayHigh}2000 & \cellcolor{grayMed}300 & \cellcolor{gpuLow}288   & \cellcolor{hgpuMed}300 & \cellcolor{energyMed}12.4  & \cellcolor{costMed}64.8 \\
\cellcolor{grayHigh}2000 & \cellcolor{grayHigh}400 & \cellcolor{gpuLow}50    & \cellcolor{hgpuHigh}400 & \cellcolor{energyLow}11.1  & \cellcolor{costMed}70.6 \\
\cellcolor{grayHigh}2000 & \cellcolor{grayHigh}500 & \cellcolor{gpuLow}0     & \cellcolor{hgpuHigh}495 & \cellcolor{energyLow}11.0  & \cellcolor{costHigh}75.4 \\
\bottomrule
\end{tabular}
\caption{\sysname{} serving multiple workflows varying resource constraints minimizing energy.
As more H100 GPUs are available, \sysname{} trades off cost for energy (same SLO).
}
\vspace{-10pt}
\label{tab:changing_resources}
\end{table}

\begin{figure}
    \includegraphics[width=0.49\textwidth]{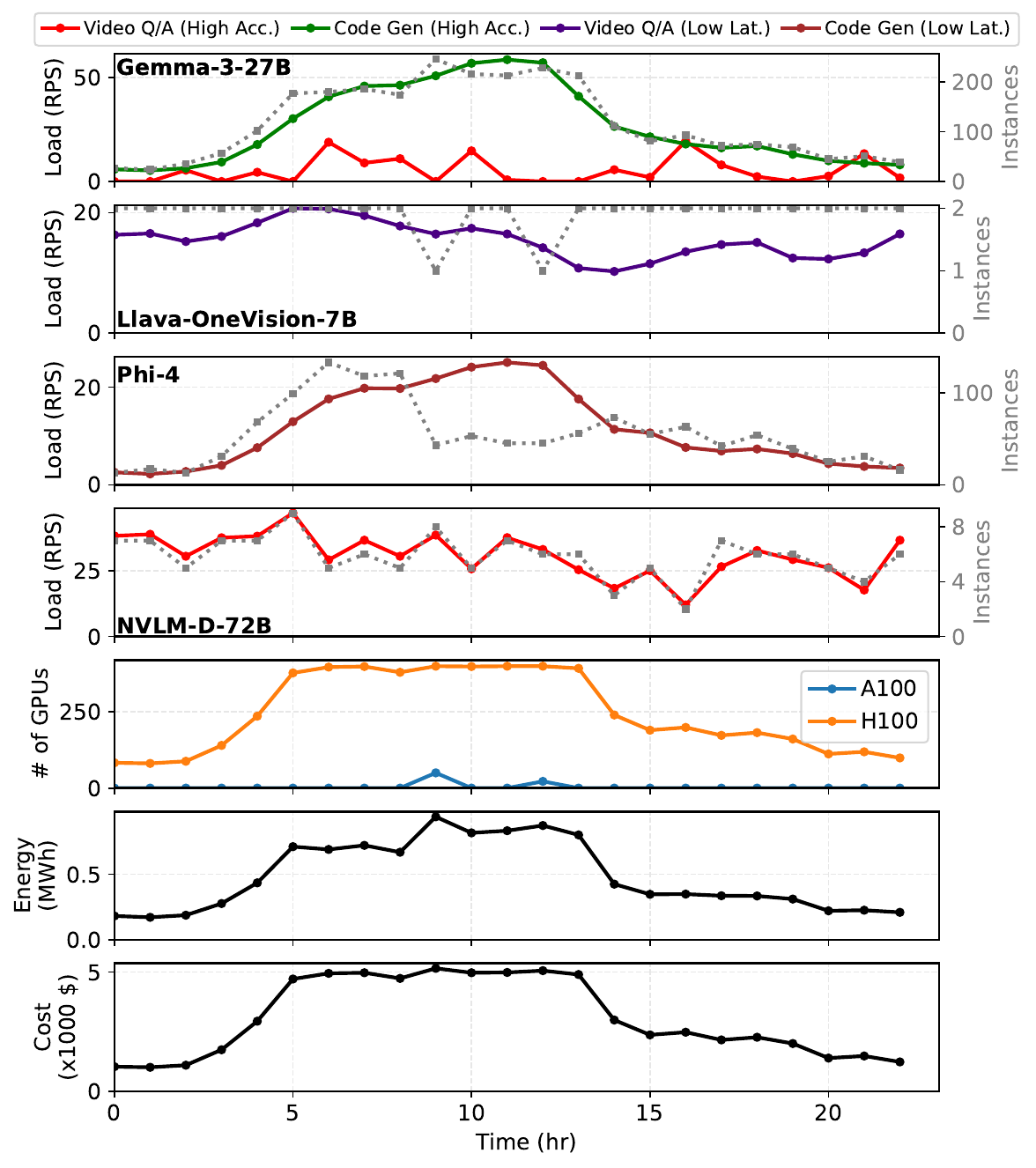}
    \caption{\sysname{} adjusts resource allocation and model instances across all workflow-SLO combinations with changing load (under a constraint of 400$\times$H100 GPUs.)}
    \vspace{-8pt}
    \label{fig:changing-resources-h100-400}
\end{figure}

\begin{figure*}[t]
  \centering
  \begin{subfigure}[b]{0.32\linewidth}
    \includegraphics[width=\textwidth]{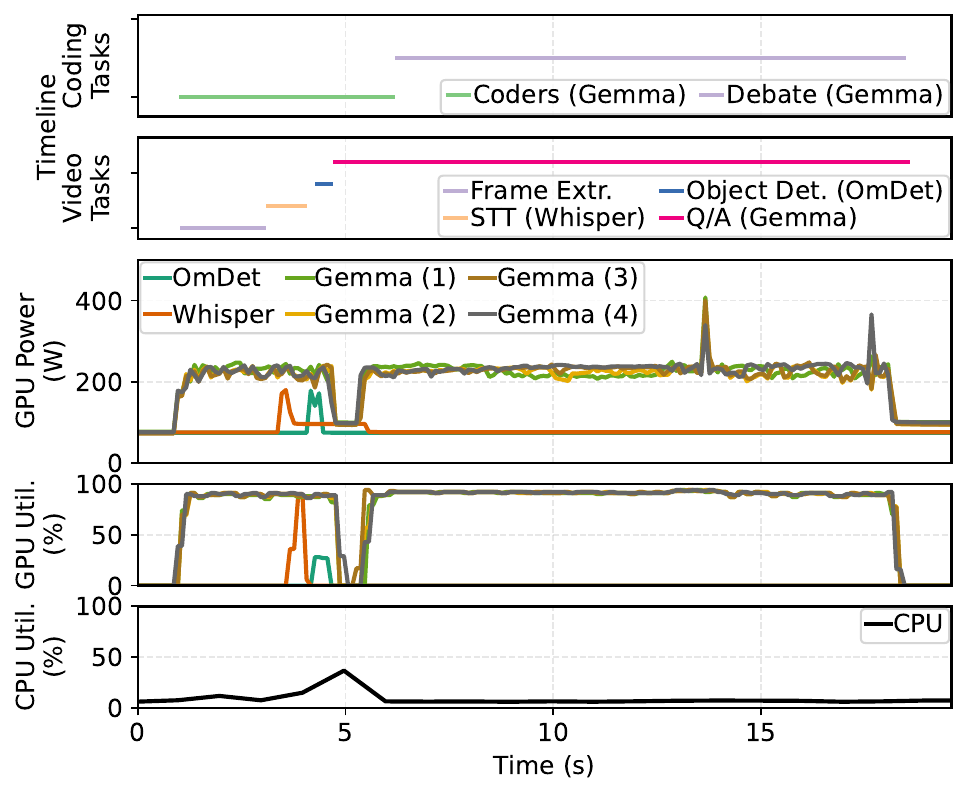}
    \caption{OmDet on 1$\times$A100, Whisper on 1$\times$A100, Gemma on 4$\times$A100s. End-to-end execution completes \textit{within} latency SLO \textit{(using 6 GPUs)}.}
    \vspace{-8pt}
    \label{fig:ovd-gpu-stt-gpu}
  \end{subfigure}\hfill
  \begin{subfigure}[b]{0.32\linewidth}
    \includegraphics[width=\textwidth]{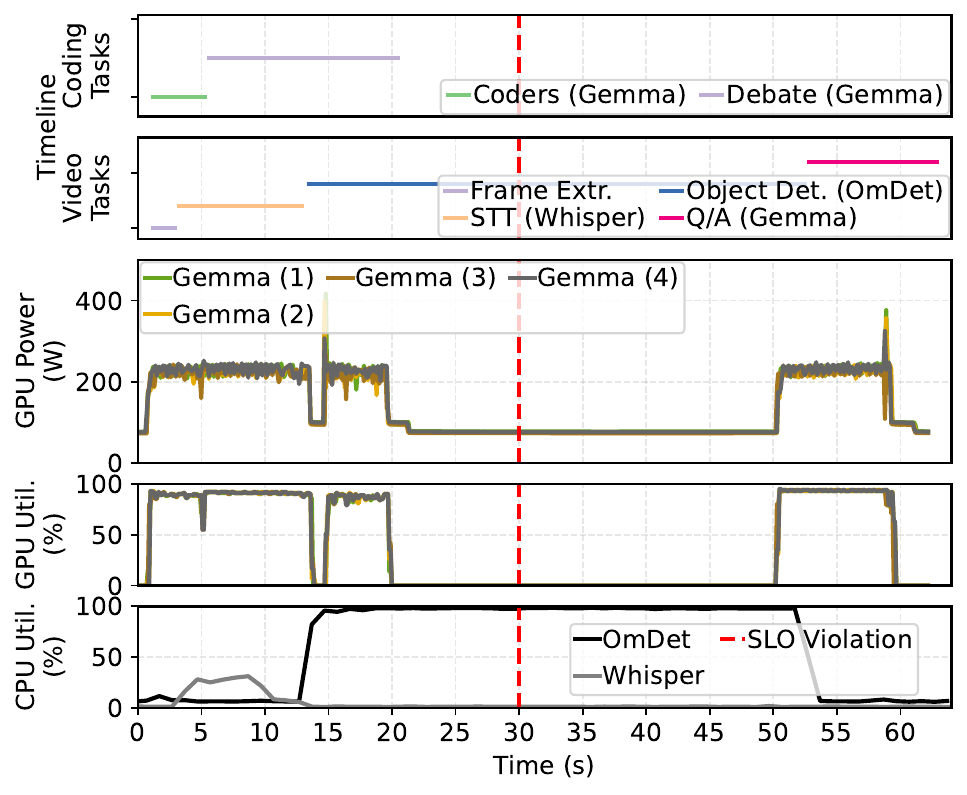}
    \caption{OmDet on CPUs, Whisper on CPUs, Gemma on 4$\times$A100s. End-to-end execution \textit{\textbf{violates}} latency SLO \textit{(using 4 GPUs)}.}
    \vspace{-8pt}
    \label{fig:ovd-cpu-stt-cpu}
  \end{subfigure}\hfill
  \begin{subfigure}[b]{0.32\linewidth}
    \includegraphics[width=\textwidth]{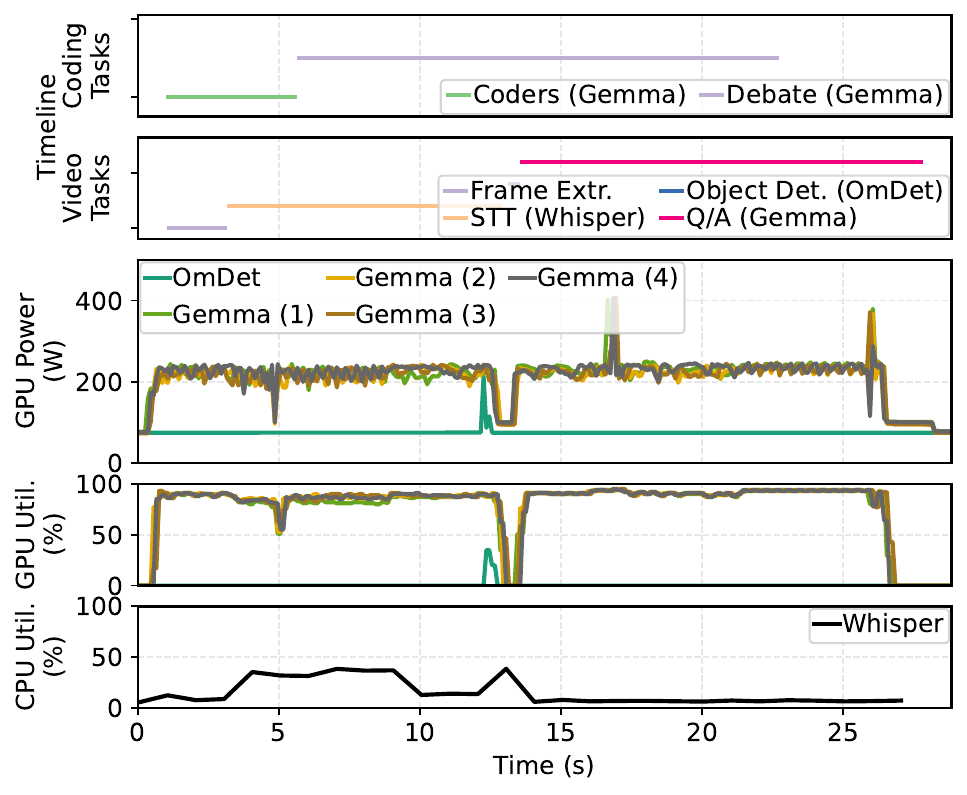}
    \caption{OmDet on 1$\times$A100, Whisper on CPUs, Gemma on 4$\times$A100s. End-to-end execution completes \textit{within} latency SLO \textit{(using 5 GPUs)}.}
    \vspace{-8pt}
    \label{fig:ovd-gpu-stt-cpu-1x}
  \end{subfigure}
  \caption{Executing a user request involving parallel video Q/A and code generation.
  \sysname{} selects the configuration in \Cref{fig:ovd-gpu-stt-cpu-1x} to minimize energy use while meeting the 30-second latency SLO and maintaining response quality.}
  \label{fig:multi-workflow-timelines}
\end{figure*}

To evaluate how \sysname{} adapts to changes in resources (\eg{} spot-instances, cloud provider prioritizing resources for other services), we run the same arrival pattern and workflow-SLO distribution as in \Cref{sec:eval:multi}.
However, we constrain the type and amount of resources available to \sysname{}.
We assume a fixed number of A100 GPUs available in the cluster (2000) and vary the number of H100 GPUs available from 0 to 500, increasing by 100 each time.
We let \sysname{} optimize to minimize energy consumption when serving the incoming workflow requests.

\Cref{tab:changing_resources} shows that \sysname{} leverages the H100 GPUs as they become available, while reducing the number of A100 GPUs.
Starting with an allocation of 1292 A100 GPUs and no H100 GPUs, which consumes 24.7 MWh energy, \sysname{} adapts to the changing H100 availability and leverage most of them (up to 495 H100s) to bring down the energy consumption to 11 MWh.
This shows that \sysname{} adjusts its resource allocation under resource-constrained settings to satisfy the SLOs while striving for higher resource efficiency.

\myparagraph{Execution Analysis}
\Cref{fig:changing-resources-h100-400} shows \sysname{} adapting to changes in load.
We consider the \emph{400$\times$H100 GPUs} configuration (without loss of generality) and observe that:
\begin{enumerate}[leftmargin=*]
    \item \sysname{} \emph{scales model instances up/down} after every optimization epoch to adapt to the shift in load dynamics.
    \item \sysname{} \emph{changes GPU allocation} with system load.
    It prefers H100 GPUs due to their energy efficiency and uses A100 GPUs for the rest, depending on load fluctuations (\eg{} at T=9 hours).
    \item \sysname{} \emph{reconfigures workflows to maximize colocation}.
    For example, Gemma serves code generation requests with high-accuracy SLOs and multiplexes load from video Q/A requests with high-accuracy SLOs when there is surplus capacity (\eg{} from T=5--9 hours and T=15--19).
\end{enumerate}

\subsection{Workflow/DAG-Aware Scheduling}
\label{sec:eval:aware}

Consider the request: \emph{verify the student’s coding solution from a video}, with a \emph{30-second latency SLO}.
The orchestrator constructs a DAG with a fan-out for the two sub-tasks that can execute in parallel:
(1) video Q/A (\Cref{fig:video-workflow}) to extract the student’s solution, and
(2) code generation (\Cref{fig:code-gen-workflow}) to produce a reference solution for comparison.

\myparagraph{Scheduling}
Leveraging workflow and model profiles, \sysname{} identifies viable scheduling options based on cluster resource availability (\Cref{fig:multi-workflow-timelines}).
All configurations use Gemma-3-27B on 4$\times$A100 GPUs.
We highlight setups that leverage both GPUs and CPUs for improved efficiency.

\myparagraph{OmDet and Whisper on GPUs}
\Cref{fig:ovd-gpu-stt-gpu} shows OmDet and Whisper running on dedicated A100 GPUs, with a total of 6$\times$A100 GPUs.
The two sub-tasks run in near-perfect parallel, with full overlap in execution.
However, despite meeting the latency goal, dedicated GPUs are underutilized, and CPUs are mostly idle.

\myparagraph{OmDet and Whisper on CPUs}
\Cref{fig:ovd-cpu-stt-cpu} shows Whisper and OmDet running on CPUs, reducing GPU usage to 4$\times$A100 and utilizing idle CPU resources.
Whisper runs efficiently without saturating CPUs, making it a good candidate for offloading.
In contrast, OmDet fully loads all cores and significantly increases latency, resulting in an SLO violation.

\myparagraph{OmDet on GPU, Whisper on CPU}
\Cref{fig:ovd-gpu-stt-cpu-1x} shows OmDet running on 1$\times$A100 GPU and Whisper on CPUs, with a total of 5$\times$A100 GPUs provisioned.
Both sub-tasks complete nearly simultaneously, and Whisper’s added CPU latency has minimal impact on end-to-end time.
This configuration meets the latency SLO while reducing GPU usage compared to the all-GPU setup.

\myparagraph{\sysname{} in Action}
Leveraging workflow stage visibility and pre-generated execution profiles, \sysname{} selects the configuration with OmDet on GPU and Whisper on CPU (\Cref{fig:ovd-gpu-stt-cpu-1x}) as the most resource-efficient option (balancing overall utilization and latency).
In contrast, traditional systems treat agentic workflows as opaque, lacking coordination between orchestration and scheduling.
This leads to inefficient execution, higher costs, and SLO violations.

\begin{figure}[t]
  \centering
  \begin{subfigure}[b]{0.49\linewidth}
    \includegraphics[width=\linewidth]{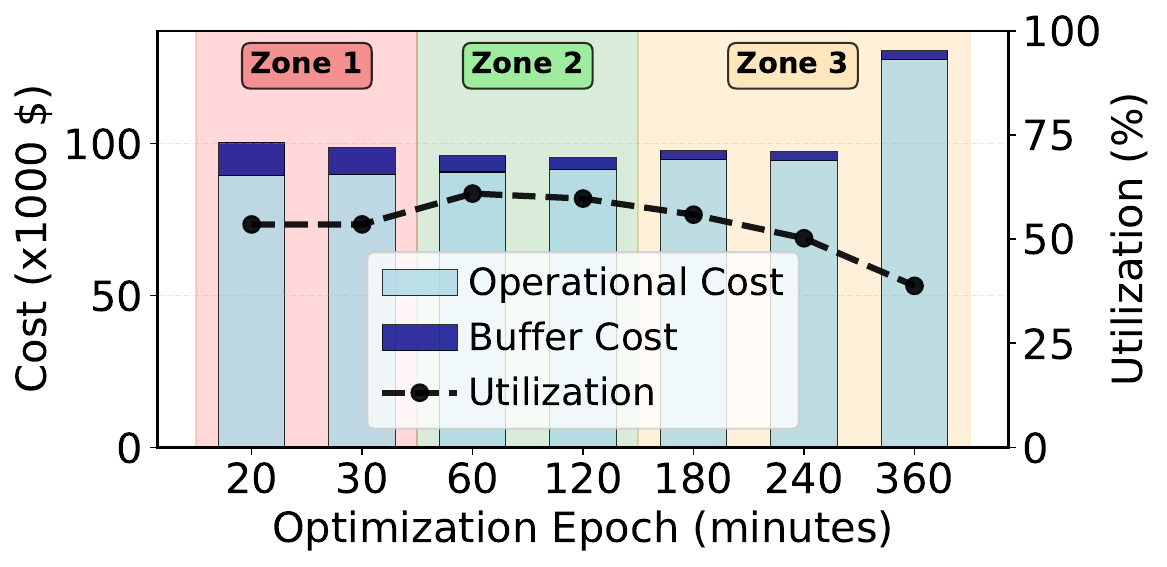}
    \caption{Cost \vs{} resource utilization.}
    \vspace{-8pt}
    \label{fig:opt_sensitivity_three_zones}
  \end{subfigure}\hfill
  \begin{subfigure}[b]{0.49\linewidth}
    \includegraphics[width=\linewidth]{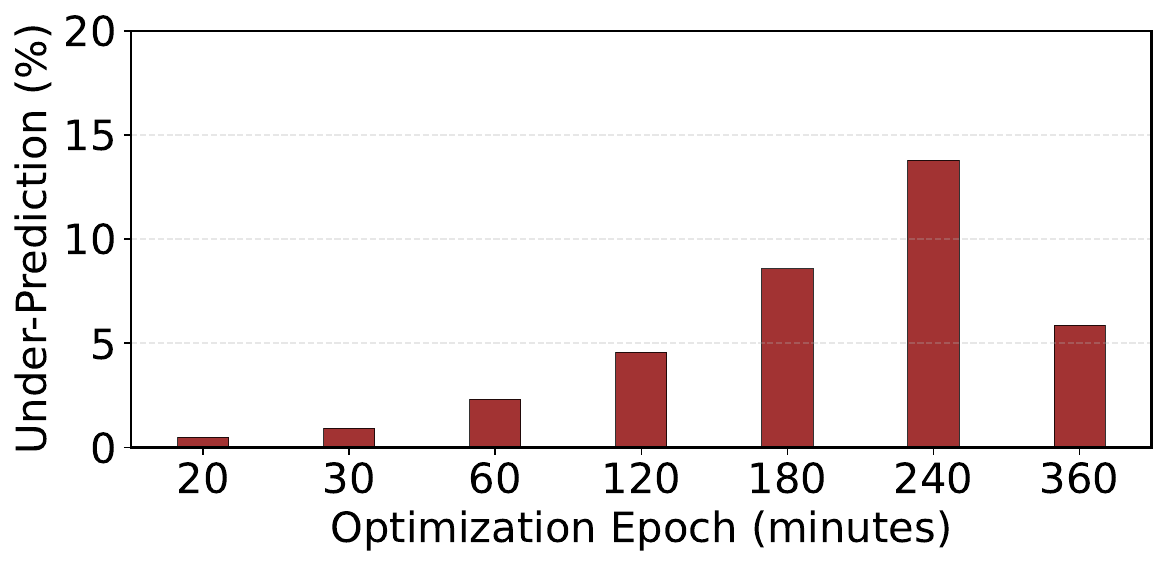}
    \caption{Demand under-prediction.}
    \vspace{-8pt}
    \label{fig:opt_sensitivity_shortage_risk}
  \end{subfigure}
  \caption{\sysname{} sensitivity to optimization epoch.}
  \label{fig:opt_sensitivity}
\end{figure}

\subsection{Optimization Frequency Sensitivity Analysis}\label{sec:eval:sensitivity}
The choice of optimization epoch presents a fundamental trade-off between resource efficiency, cost, and system responsiveness.
We evaluate intervals from 20 minutes to 6 hours, measuring cost, utilization, and worst-case dropped requests.
Provisioning new instances (\ie{} VM allocation, software setup, and model transfer to GPUs) is assumed to take 20 minutes~\cite{fu2024serverlessllm,hao2021empirical,doca}.
We use an exponentially weighted moving average (EWMA)~\cite{cisar2010ewma}, with $\alpha=0.5$, to predict workload demand at every epoch.

\myparagraph{Three-Zone Cost Structure}
\Cref{fig:opt_sensitivity_three_zones} shows three distinct zones that guide optimal selection.

\myparagraphemph{Zone 1 (10--60 minutes): Buffer-dominated}
Frequent reoptimization induces high transition overhead.
Excessive GPU provisioning during transitions leads to lower utilization, despite responsive demand adaptation.
Frequent model and tool changes can also reduce KV cache efficiency for LLMs.

\myparagraphemph{Zone 2 (60--180 minutes): Balanced}
Transition costs and prediction uncertainty offset, yielding peak cost efficiency.
Utilization peaks at an epoch of around 60 minutes, reflecting the best balance between adaptation frequency and stability.

\myparagraphemph{Zone 3 (180+ minutes): Uncertainty-dominated}
At long intervals, buffer cost from transition overhead is minimal, but coarse-grained provisioning increasingly diverges from fine-grained, dynamic demand. This reduces prediction accuracy and drives over-provisioning, lowering efficiency.

\myparagraph{Worst-case System Responsiveness}
We measure responsiveness by demand under-prediction (requests that would be dropped without auto-scaling).
This reflects the \emph{worst-case}, as auto-scalers can mitigate mismatches post-provisioning.
\Cref{fig:opt_sensitivity_shortage_risk} shows under-prediction is minimal at short intervals but rises steadily, peaking at $\sim$15\% around 240 minutes.
Beyond that, longer intervals lead to over-provisioning, lowering under-prediction but hurting utilization.

\myparagraph{Operational Implications}
This analysis reveals a key trade-off: short intervals improve responsiveness but increase transition costs, while long intervals reduce overhead but hurt prediction accuracy and utilization.
Providers can tune this balance based on priorities (\eg{} cost, utilization, responsiveness).
In our setup, a 60-minute interval offers a middle ground (high utilization, low cost, and manageable shortage risk).
While we use EWMA for forecasting, the trade-off holds with more advanced predictors as well.
\section{Related Work}

\myparagraph{Agentic Workflow Development}
Frameworks like LangGraph~\cite{langgraph}, LangChain~\cite{langchain}, and AutoGen~\cite{autogen} build agentic workflows \textit{imperatively}, composing model and tool calls.
DSPy~\cite{dspy} and Palimpzest~\cite{liu2024declarative,palimpzestCIDR} take a \textit{declarative} approach, focusing on prompt and query optimization.
However, these frameworks still blur configuration and logic, burden developers with resource management, and struggle to scale efficiently across large configuration spaces.

\myparagraph{Automated Workflow Generation}
There has been extensive work from the ML community on automating workflow generation to improve response quality~\cite{liu2025divideoptimizemergefinegrained,niu2025flow,li2024autoflow,zhang2024aflow,wu2025optimas,liu2024declarative}.
This line of work is complementary to \sysname{}, which can integrate these workflow generation techniques into the \sysname{} workflow orchestrator (\Cref{sec:design:workflow-specification}).

\myparagraph{Systems Optimization}
Prior systems have focused on accelerating workflow execution with improved data flow management~\cite{raghavan2025altoorchestratingdistributedcompound,tan2025teolaendtoendoptimizationllmbased} and scheduling of LLM inference calls in the context of agentic applications~\cite{luo2025autellixefficientservingengine,lin2024parrot}.
Others target response quality by exploring model selection~\cite{chen2025optimizingmodelselectioncompound,feng2025graphroutergraphbasedrouterllm} and scaling properties~\cite{chen2024llmcallsneedscaling,ong2025routellmlearningroutellms} of test-time compute.
Systems like SpotServe~\cite{miao2023spotserveservinggenerativelarge} and Loki~\cite{ahmad2024loki} address resource and load dynamism, but remain confined to single-model serving.
Recent works~\cite{jin2025energycostreasoninganalyzing,kim2025costdynamicreasoningdemystifying,chung2025mlenergybenchmarkautomatedinference} have examined the energy costs of test-time compute and their impact on response quality. In contrast, \sysname{} introduces a fully automated system to navigate these complex tradeoffs for agentic workflows.
\section{Conclusion}

\sysname{} brings a declarative programming model and adaptive runtime to the serving of multi-tenant agentic workflows.
By decoupling workflow logic from execution configurations and tightly integrating orchestration with cluster management, \sysname{} enables dynamic reconfiguration and resource-aware scheduling while honoring diverse SLOs.
Our evaluation shows that \sysname{} substantially improves efficiency (reducing GPU usage, energy consumption, and cost) without compromising workflow quality or latency.
We see opportunities in extending adaptive co-optimization across larger clusters, supporting more heterogeneous accelerators, and exploring workload specialization for emerging classes of agentic applications.

\bibliographystyle{ACM-Reference-Format}
\bibliography{paper}

\clearpage
\appendix
\section{Appendix}\label{sec:appendix}

\subsection{Math Q/A}\label{sec:appendix:math}

\begin{figure}
    \includegraphics[width=0.49\textwidth]{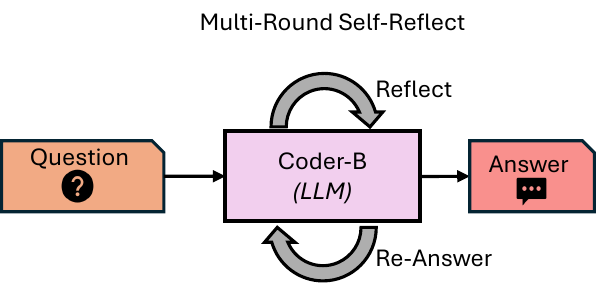}
    \caption{Math Q/A workflow (self-reflect~\cite{renze2024self} structure).}
    \label{fig:math-workflow}
\end{figure}

Math Q/A, as shown in \Cref{fig:math-workflow}, is a workflow to answer mathematical problems.
It uses a self-reflect~\cite{renze2024self} structure where a \emph{mathematician} agent generates a response to the question, self-reflects on the response, and iteratively continues this until it is confident in the answer or a limit on the maximum number of rounds is reached.

\subsubsection{Characterization}
\begin{figure*}[t]
    \centering
    \begin{subfigure}[b]{0.33\textwidth}
        \includegraphics[width=\textwidth]{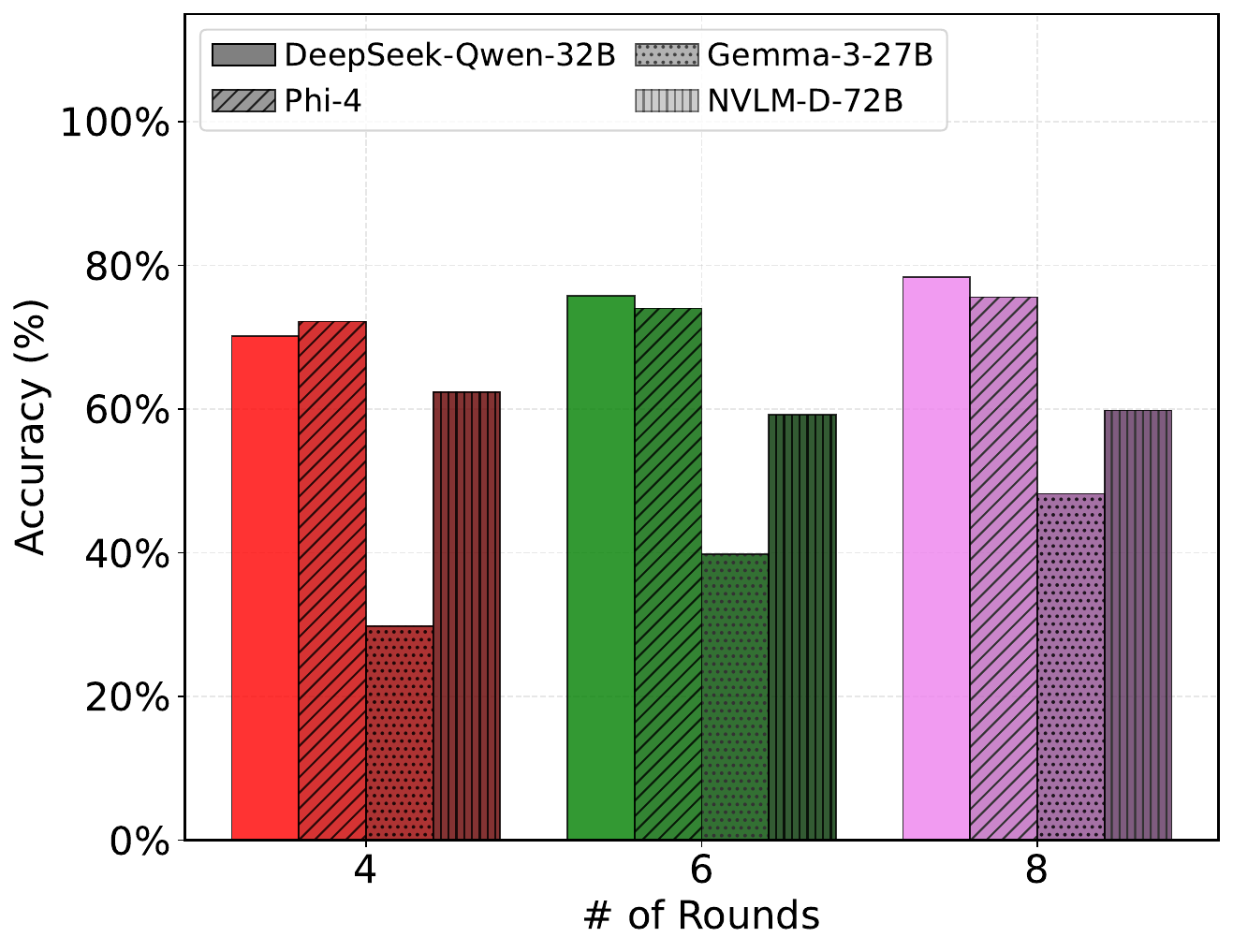}
        \caption{Math Q/A accuracy.}
        \label{fig:math-accuracy}
    \end{subfigure}\hfill
    \begin{subfigure}[b]{0.33\textwidth}
        \includegraphics[width=\textwidth]{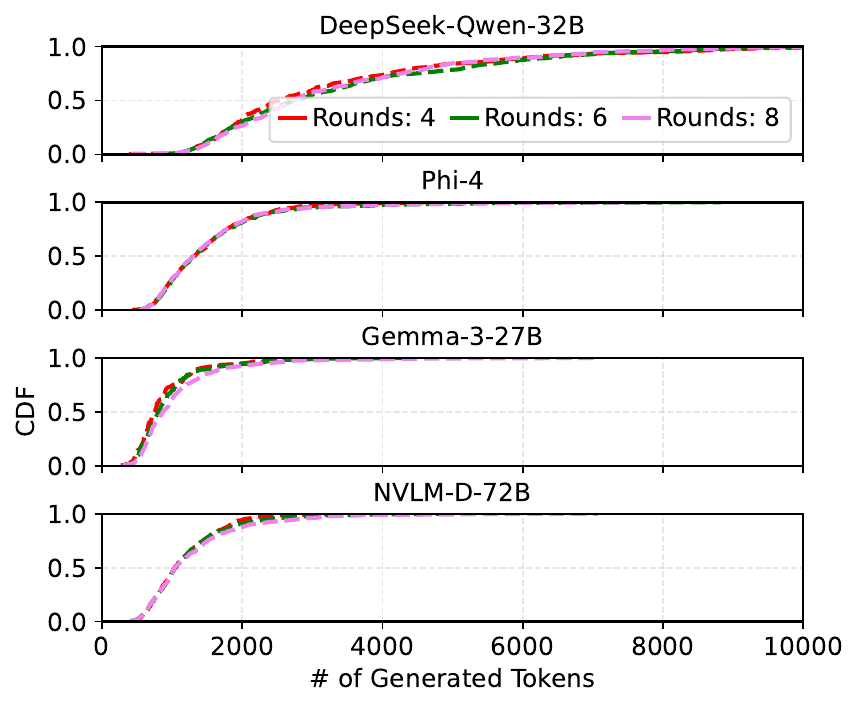}
        \caption{Math Q/A model load (generated tokens).}
        \label{fig:math-num-completion-tokens}
    \end{subfigure}\hfill
    \begin{subfigure}[b]{0.33\textwidth}
        \includegraphics[width=\textwidth]{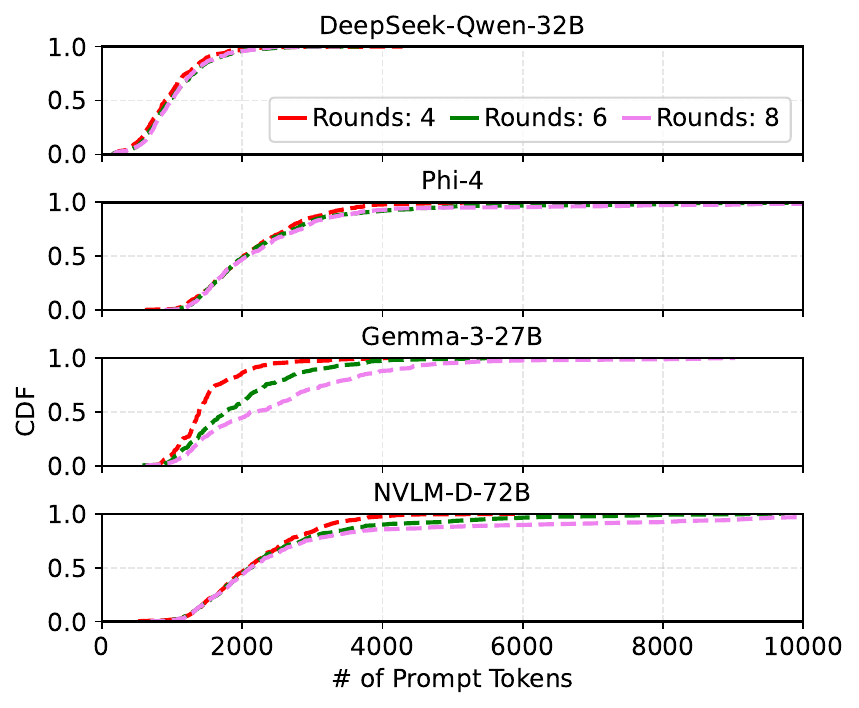}
        \caption{Math Q/A model load (prompt tokens).}
        \label{fig:math-num-prompt-tokens}
    \end{subfigure}\hfill
    \caption{Workflow accuracy under different configurations and the token processing load on the respective models.}
    \label{fig:math-workflow-profile}
\end{figure*}

We present the results (\Cref{fig:math-workflow-profile}) for different models and different number of self-refine rounds (\texttt{R}) (\Cref{fig:math-accuracy}) and the token generation load in terms of prompt (\Cref{fig:math-num-prompt-tokens}) and completion tokens (\Cref{fig:math-num-completion-tokens}) for different configurations.

\subsubsection{Single-Workflow Optimization}

\begin{figure}[t]
  \centering
  \begin{subfigure}[b]{0.49\linewidth}
    \includegraphics[width=\linewidth]{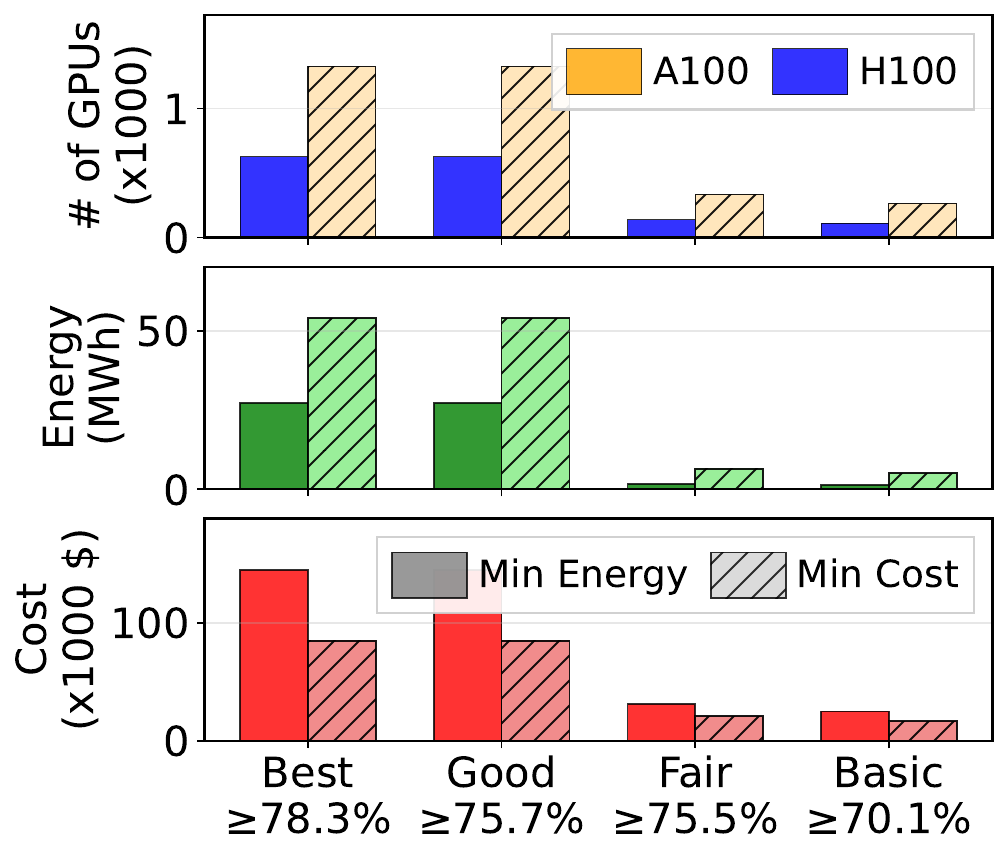}
    \caption{Accuracy SLOs.}
    \label{fig:math-accuracy-slo}
  \end{subfigure}\hfill
  \begin{subfigure}[b]{0.49\linewidth}
    \includegraphics[width=\linewidth]{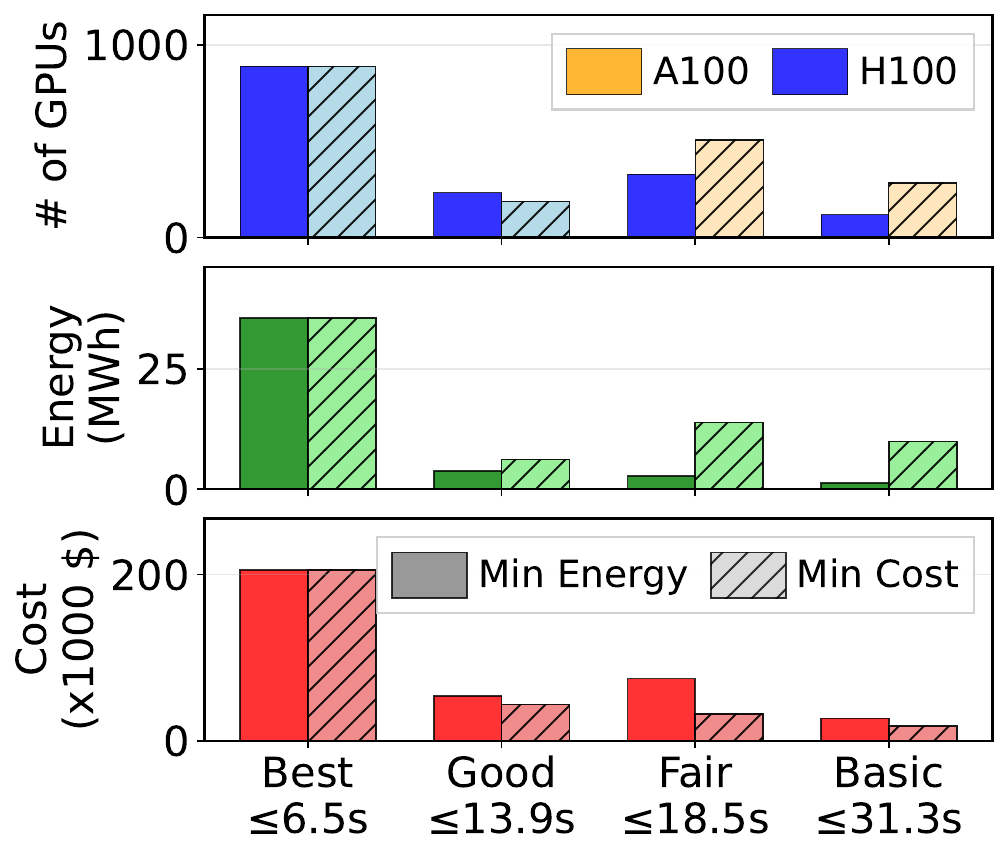}
    \caption{Latency SLOs.}
    \label{fig:math-latency-slo}
  \end{subfigure}
  \caption{Math Q/A workflow configured for different SLO and optimization objectives.}
  \label{fig:math-optimizations}
\end{figure}

Similar to \Cref{sec:eval:single}, we optimize the math Q/A workflow using \sysname{} for different accuracy and latency SLO tiers and optimization objectives.
The results are presented in \Cref{fig:math-optimizations}.

\subsubsection{Multi-Workflow Optimization}
\begin{figure*}
    \centering
    \begin{subfigure}[b]{0.49\linewidth}
        \includegraphics[width=\textwidth]{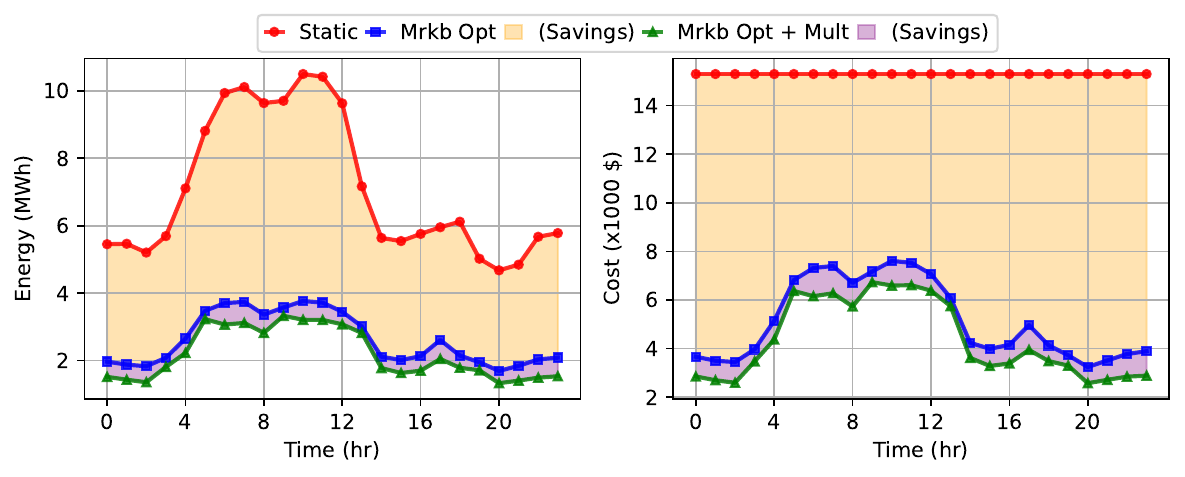}
        \caption{Energy and cost over time. Static policy is agnostic to load shifts while \sysname{} reconfigures workflows for higher efficiency.}
        \vspace{-8pt}
        \label{fig:multiplex-math-code-gen-static-conservative-a100-compare}
    \end{subfigure}\hfill
    \begin{subfigure}[b]{0.49\linewidth}
        \includegraphics[width=\textwidth]{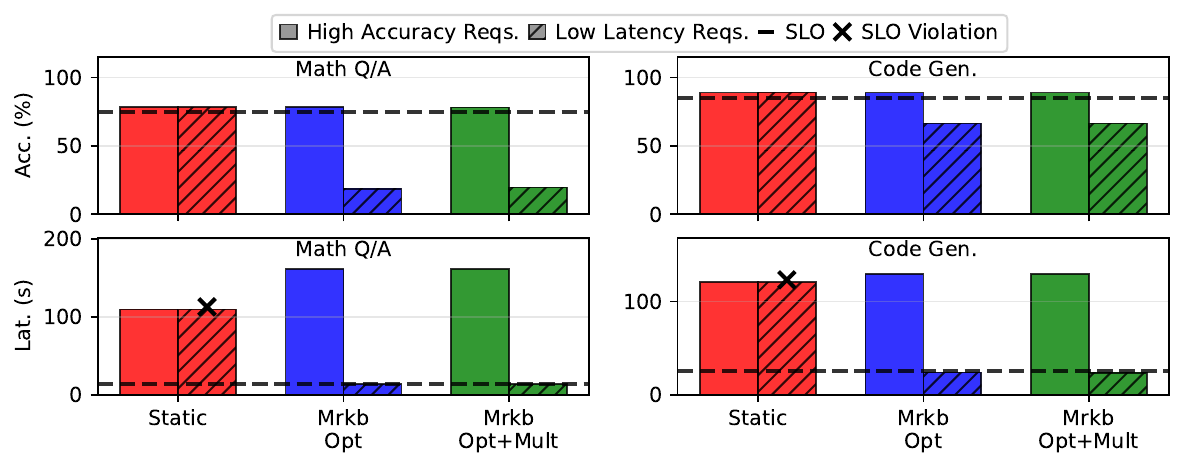}
        \caption{Accuracy and latency for all requests. Static policy does not distinguish between request SLO categories.}
        \vspace{-8pt}
        \label{fig:multiplex-math-code-gen-static-conservative-a100-normalized-comparison}
    \end{subfigure}
    \caption{Comparing a hand-crafted static configuration (only A100s) to: (1) \sysname{} optimizing individual workflows, and (2) \sysname{} jointly-optimizing across workflows + multiplexing resources (Math Q/A + Code Gen).}
    \label{fig:multiplex-math-code-gen-static-conservative-a100}
\end{figure*}

\begin{table}[t]
\centering
\scriptsize
\setlength{\tabcolsep}{3pt}
\begin{tabular}{c c c c}
\toprule
\textbf{Policy} &
\makecell{\textbf{\# of GPUs}} &
\makecell{\textbf{Energy}\\\textbf{(MWh)}} &
\makecell{\textbf{Cost}\\\textbf{($\times$1000 \$)}} \\
\midrule
Static        & 4448 & 169.87 & 367.2 \\
\sysnameshort{} Opt      & 1875 & 62.88  & 123.0 \\
\sysnameshort{} Opt+Mult & 1660 & 52.66  & 104.6 \\
\bottomrule
\end{tabular}
\caption{Comparison of resource usage, energy, and cost. \sysname{} substantially reduces GPUs, energy, and cost compared to static allocation, with further gains from multiplexing across all math Q/A and code generation workflow requests.}
\label{tab:multiplex-math-code-gen-static-conservative-a100-absolute}
\end{table}

Similar to \Cref{sec:eval:multi}, we evaluate end-to-end execution of running 24 hours of Azure traces~\cite{stojkovic2025dynamollm}.
This time, the two categories of requests from the traces map to two agentic workflows: \emph{chat} requests maps to \emph{math Q/A} workflow requests and \emph{coding} requests map to the \emph{code generation} workflow (\Cref{fig:code-gen-workflow}).
The results are shown in \Cref{fig:multiplex-math-code-gen-static-conservative-a100}.
The hand-crafted \emph{static} baseline policy (similar to LangGraph~\cite{langgraph}) has a fixed allocation of DeepSeek-Qwen-32B for the math Q/A workflow and Gemma-3-27B for code generation workflow -- balancing quality and resource usage.
As explained in \Cref{sec:eval}, the static baseline has no visibility into the workflow, load or resource availability and suffers from under-utilization.
\sysname{} on the other hand is adaptive to load for each workflow--SLO combination and appropriately configures workflows at each optimization epoch, resulting in massive resource savings.
When allowed to multiplex across resources, it yields further savings by colocating the requests from all workflows onto shared model instances whenever appropriate.
\Cref{tab:multiplex-math-code-gen-static-conservative-a100-absolute} summarizes the results for the 24 hour duration.
We observe a reduction of $\approx 2.7\times$ in GPUs, $\approx3.2\times$ in energy, and $\approx3.5\times$ in cost from the static baseline to \sysname{}'s best optimization policy.

\subsection{Video}\label{sec:appendix:video}

\begin{table}[t]
\centering
\tiny
\setlength{\tabcolsep}{1.5pt}
\begin{tabular}{p{0.4cm}p{0.8cm}p{0.6cm}p{1.9cm}cp{0.8cm}p{0.2cm}p{0.8cm}p{0.8cm}}
\toprule
SLO & Objective & Tier & Model & Frames & GPU & TP & TPOT (s) & TPS \\
\midrule
\multirow{8}{*}{Acc.} & \multirow{4}{*}{Cost} & Best & Gemma-3-27B & 10 & A100 & 4 & 0.0624 & 699 \\
 &  & Fair & NVLM-D-72B & 5 & A100 & 4 & 0.0966 & 325 \\
 &  & Basic & Llava-OneVision-7B & 5 & A100 & 4 & 0.0224 & 2244 \\
 &  & Good & Gemma-3-27B & 5 & A100 & 4 & 0.0624 & 700 \\
\cmidrule(lr){2-9}
 & \multirow{4}{*}{Energy} & Best & Gemma-3-27B & 10 & H100 & 4 & 0.0484 & 1688 \\
 &  & Fair & NVLM-D-72B & 5 & H100 & 4 & 0.0650 & 766 \\
 &  & Basic & Llava-OneVision-7B & 5 & H100 & 4 & 0.0079 & 3271 \\
 &  & Good & Gemma-3-27B & 5 & H100 & 4 & 0.0472 & 1668 \\
\midrule
\multirow{8}{*}{Lat.} & \multirow{4}{*}{Cost} & Best & Llava-OneVision-7B & 1 & H100 & 4 & 0.0044 & 479 \\
 &  & Fair & Llava-OneVision-7B & 1 & A100 & 4 & 0.0224 & 2244 \\
 &  & Basic & Llava-OneVision-7B & 1 & A100 & 4 & 0.0224 & 2244 \\
 &  & Good & Llava-OneVision-7B & 1 & A100 & 4 & 0.0085 & 926 \\
\cmidrule(lr){2-9}
 & \multirow{4}{*}{Energy} & Best & Llava-OneVision-7B & 1 & H100 & 4 & 0.0044 & 479 \\
 &  & Fair & Llava-OneVision-7B & 1 & H100 & 4 & 0.0070 & 2836 \\
 &  & Basic & Llava-OneVision-7B & 1 & H100 & 4 & 0.0070 & 2836 \\
 &  & Good & Llava-OneVision-7B & 1 & H100 & 4 & 0.0070 & 2836 \\
\bottomrule
\end{tabular}
\caption{Video Q/A workflow configurations chosen by \sysname{} corresponding to the experiment in \Cref{sec:eval:single}}
\label{tab:video_configs_gpu}
\end{table}

We present the most commonly chosen configuration by \sysname{} for each workflow--SLO combination of the video Q/A workflow under different optimization objectives in \Cref{tab:video_configs_gpu} (corresponding to the experiment presented in \Cref{sec:eval:single}).
It shows various knobs (\eg{} model, number of frames, GPU type \etc) and the offered load per model instance in tokens-per-second (TPS) with the observed latency in time-per-output-token (TPOT)
For example, we can observe the changing model and number of frames processed as the accuracy SLO is relaxed.
For the latency SLO requests, \sysname{} keeps the same workflow- and model-level knobs but changes GPU type and increases the allowed load per model instance to increase batching as the SLO is relaxed.

\subsection{Code Generation}\label{sec:appendix:code-gen}

\begin{table}[t]
\centering
\tiny
\setlength{\tabcolsep}{1.5pt}
\begin{tabular}{p{0.4cm}p{0.8cm}p{0.6cm}p{1.9cm}ccp{0.8cm}p{0.2cm}p{0.8cm}p{0.8cm}}
\toprule
SLO & Objective & Tier & Model & Debaters & Rounds & GPU & TP & TPOT (s) & TPS \\
\midrule
\multirow{8}{*}{Acc.} & \multirow{4}{*}{Cost} & Best & DeepSeek-Qwen-32B & 4 & 4 & A100 & 4 & 0.0767 & 653 \\
 &  & Good & Gemma-3-27B & 4 & 4 & A100 & 4 & 0.0624 & 700 \\
 &  & Fair & Gemma-3-27B & 2 & 4 & A100 & 4 & 0.0624 & 700 \\
 &  & Basic & Phi-4 & 2 & 4 & A100 & 2 & 0.0609 & 623 \\
\cmidrule(lr){2-10}
 & \multirow{4}{*}{Energy} & Best & DeepSeek-Qwen-32B & 4 & 4 & H100 & 4 & 0.0387 & 1390 \\
 &  & Good & Gemma-3-27B & 4 & 4 & H100 & 4 & 0.0496 & 1709 \\
 &  & Fair & Gemma-3-27B & 2 & 4 & H100 & 4 & 0.0487 & 1693 \\
 &  & Basic & Phi-4 & 2 & 4 & H100 & 1 & 0.0373 & 757 \\
\midrule
\multirow{8}{*}{Lat.} & \multirow{4}{*}{Cost} & Best & NVLM-D-72B & 2 & 2 & H100 & 8 & 0.0129 & 84 \\
 &  & Good & Phi-4 & 2 & 2 & H100 & 2 & 0.0169 & 1036 \\
 &  & Fair & Phi-4 & 2 & 2 & H100 & 2 & 0.0218 & 1185 \\
 &  & Basic & Phi-4 & 2 & 2 & A100 & 2 & 0.0609 & 623 \\
\cmidrule(lr){2-10}
 & \multirow{4}{*}{Energy} & Best & NVLM-D-72B & 2 & 2 & H100 & 8 & 0.0129 & 84 \\
 &  & Good & Phi-4 & 2 & 2 & H100 & 1 & 0.0165 & 248 \\
 &  & Fair & Phi-4 & 2 & 2 & H100 & 1 & 0.0253 & 552 \\
 &  & Basic & Phi-4 & 2 & 2 & H100 & 1 & 0.0372 & 755 \\
\bottomrule
\end{tabular}
\caption{Code generation workflow configurations chosen by \sysname{} corresponding to the experiment in \Cref{sec:eval:single}}
\label{tab:code_generation_configs_gpu}
\end{table}

We present the most commonly chosen configuration by \sysname{} for each workflow--SLO combination of the code generation workflow under different optimization objectives in \Cref{tab:code_generation_configs_gpu} (corresponding to the experiment presented in \Cref{sec:eval:single}).
It shows various knobs (\eg{} model, number of debaters, number of debate rounds \etc) and the offered load per model instance in tokens-per-second (TPS) with the observed latency in time-per-output-token (TPOT)
For example, we can observe the changing model and number of debaters as the accuracy SLO is relaxed.
For the latency SLO requests, \sysname{} mostly keeps the same workflow- and model-level knobs but changes GPU type and increases the allowed load per model instance to increase batching as the SLO is relaxed.
Notably, it also changes the tensor parallelism for the same model (Phi-4~\cite{abdin2024phi4technicalreport}) between optimizing for cost and energy.

\subsection{Azure Traces}\label{sec:appendix:traces}
\begin{figure}
    \includegraphics[width=0.49\textwidth]{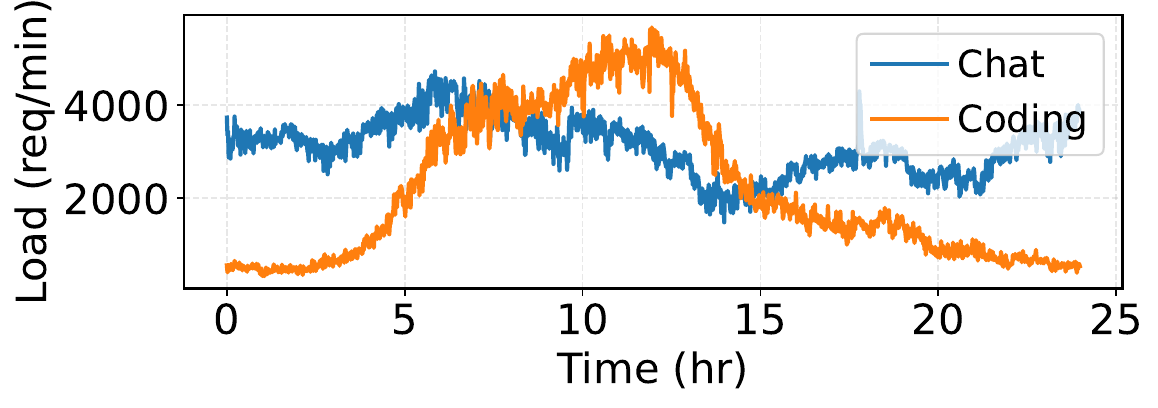}
    \caption{Azure LLM serving traces.}
    \vspace{-8pt}
    \label{fig:azure_arrivals}
\end{figure}

We use a subset of LLM serving traces released by Azure~\cite{stojkovic2025dynamollm} from 08:00 05/15/2024 to 08:00 05/16/2024 shown in \Cref{fig:azure_arrivals}.

\subsection{Optimization Formulation}\label{sec:optimization-milp}
We formulate our optimization problem (\Cref{sec:design:optimization}) as a temporal resource allocation task for serving multiple workflows with heterogeneous SLOs across diverse model profiles.

\myparagraph{Sets and Indices}
\begin{itemize}[leftmargin=*,topsep=0pt,itemsep=0pt]
    \item $\mathcal{W}$: workflows
    \item $\mathcal{S}$: SLO types
    \item $\mathcal{M}$: model profiles
    \item $\mathcal{C}_{w}$: workflow configurations for $w$
    \item $\mathcal{G}$: resource types
\end{itemize}

\myparagraph{Parameters}
\begin{itemize}[leftmargin=*,topsep=0pt,itemsep=0pt]
    \item $\lambda_{w,s}^{\text{peak}}$: Peak request rate for workflow $w$ with SLO $s$
    \item $\lambda_{w,s}^{\text{avg}}$: Average request rate for workflow $w$ with SLO $s$
    \item $\alpha$: Unified buffer factor (default 1.15)
    \item $\tau_{w,s}$: SLO threshold for workflow $w$ and SLO type $s$
    \item $a_{c}$: Accuracy of workflow configuration $c \in \mathcal{C}_w$
    \item $t_{c}$: Tokens per request for workflow configuration $c$
    \item $\theta_m$: Token throughput (tokens/sec) for model profile $m$
    \item $\ell_m^{\text{TTFT}}$: Time to first token for model profile $m$
    \item $\ell_m^{\text{TPOT}}$: Time per output token for model profile $m$
    \item $g_m$: Parallelism for model $m$
    \item $e_m$: Energy consumption (kWh) for model profile $m$
    \item $c_g$: Cost per instance per second for resource type $g \in \mathcal{G}$
    \item $B_g$: Maximum available resource instances of type $g$
\end{itemize}

\myparagraph{Decision Variables}
\begin{itemize}[leftmargin=*,topsep=0pt,itemsep=0pt]
    \item $n_m \in \mathbb{Z}^+$: Number of instances of model profile $m$
    \item $x_{w,s,c,m}^{\text{peak}} \in \mathbb{R}^+$: Peak load allocation from $(w,s,c)$ to model $m$
    \item $x_{w,s,c,m}^{\text{avg}} \in \mathbb{R}^+$: Average load allocation from $(w,s,c)$ to model $m$
\end{itemize}

\myparagraph{Constraints}

\textbf{Demand Satisfaction (Peak):}

Ensure peak demand is met with buffer:
\begin{equation}
\lambda_{w,s}^{\text{peak}} \leq \sum_{c \in \mathcal{C}_w, m \in \mathcal{M}} x_{w,s,c,m}^{\text{peak}} \leq \alpha \cdot \lambda_{w,s}^{\text{peak}}, \quad \forall w \in \mathcal{W}, s \in \mathcal{S}
\end{equation} \\

\textbf{Demand Satisfaction (Average):}

Similar bounds for average demand:
\begin{equation}
\lambda_{w,s}^{\text{avg}} \leq \sum_{c \in \mathcal{C}_w, m \in \mathcal{M}} x_{w,s,c,m}^{\text{avg}} \leq \alpha \cdot \lambda_{w,s}^{\text{avg}}, \quad \forall w \in \mathcal{W}, s \in \mathcal{S}
\end{equation} \\

\textbf{Capacity Constraint with Multiplexing:}

Account for statistical multiplexing:
\begin{equation}
\mu_m \cdot \sum_{w,s,c} x_{w,s,c,m}^{\text{peak}} \cdot t_c \leq n_m \cdot \theta_m, \quad \forall m \in \mathcal{M}
\end{equation}
\qquad where $\mu_m$ is the model-specific multiplexing factor. \\

\textbf{SLO Filtering:}

For accuracy SLO ($s = \text{max\_accuracy}$):
\begin{equation}
x_{w,s,c,m}^{\text{peak}} = 0 \quad \text{if } a_c < \tau_{w,s}
\end{equation}

For latency SLO ($s = \text{min\_latency}$):
\begin{equation}
x_{w,s,c,m}^{\text{peak}} = 0 \quad \text{if } \ell_m^{\text{TTFT}} + t_c \cdot \ell_m^{\text{TPOT}} > \tau_{w,s}
\end{equation} \\

\textbf{Cost Budget Constraint:}

If cost SLO is specified:
\begin{equation}
\sum_{w,s,c,m} x_{w,s,c,m}^{\text{avg}} \cdot \frac{t_c}{\theta_m} \cdot g_m \cdot c_{g(m)} \leq \text{Cost}_{\text{budget}}
\end{equation}
\qquad where $\text{Cost}_{\text{budget}} = \sum_{w \in \mathcal{W}} \tau_{w,\text{cost}} \cdot \sum_s \lambda_{w,s}^{\text{avg}}$. \\

\textbf{Resource Budget Constraint:}
\begin{equation}
\sum_{m: \text{GPU}(m)=g} n_m \cdot g_m \leq B_g, \quad \forall g \in \mathcal{G}
\end{equation} \\

\textbf{SLO Filtering Constraints:}
\begin{align}
x_{w,s,c,m}^{\text{peak}} &= 0 \quad \text{if } a_c < \tau_{w,s} \quad \text{(accuracy)} \\[0.5em]
x_{w,s,c,m}^{\text{peak}} &= 0 \quad \text{if } \ell_m^{\text{TTFT}} + t_c \ell_m^{\text{TPOT}} > \tau_{w,s} \quad \text{(latency)}
\end{align} \\

\textbf{Cost Budget Constraint:}
\begin{equation}
\sum_{w,s,c,m} x_{w,s,c,m}^{\text{avg}} \frac{t_c}{\theta_m} g_m c_{g(m)} \leq \sum_w \tau_{w,\text{cost}} \sum_s \lambda_{w,s}^{\text{avg}}
\end{equation} \\

\myparagraph{Objective Functions}

\textbf{Minimize Energy:}
\begin{equation}
\min \sum_m n_m e_m g_m
\end{equation}

\textbf{Minimize Cost:}
\begin{equation}    
\min \sum_m n_m g_m c_{g(m)}
\end{equation}

\textbf{Maximize Accuracy Under a Cost Budget:} 
\begin{equation}
\max \frac{\sum_{w,s,c,m} x_{w,s,c,m}^{\text{avg}} a_c}{\sum_{w,s} \lambda_{w,s}^{\text{avg}}} - \epsilon \cdot \text{Cost}_{\text{total}}
\end{equation}
\qquad where $\epsilon = 0.001$. \\

\myparagraph{Solution Method}

The formulated Mixed Integer Linear Program (MILP) is solved using Gurobi~\cite{gurobi} with a time limit of 300 seconds.
The solution yields an allocation of model instance counts $n_m^*$ and load distributions across model instances for all workflows.

\end{document}